\documentclass[prd,preprintnumbers,twocolumn,floatfix,nofootinbib,superscriptaddress]{revtex4}
\usepackage{amsmath,amssymb}
\usepackage{graphicx}

\newcommand{\tmdpdf}{\ensuremath{\tilde{F}}}
\newcommand{\trans}[1]{\ensuremath{{\bf #1}_{\rm T}}}

\newcommand{\gammae}{\ensuremath{\gamma_{\rm E}}}

\newcommand{\kzero}{\ensuremath{\langle k_T^2 \rangle_0^f}}

\newcommand{\MSbar}{\ensuremath{\overline{\text{MS}}}}

\begin{document}
\title{QCD Evolution of the Sivers Function}

\preprint{NIKHEF-2011-030}
\preprint{YITP-SB-11-37}

\author{S.~M.~Aybat}
\email{maybat@nikhef.nl}
\affiliation{Nikhef Theory Group, Science Park 105, 1098XG Amsterdam, The Netherlands}
\author{J.~C.~Collins}
\email{collins@phys.psu.edu}
\affiliation{Department of Physics, Pennsylvania State University,
   University Park, Pennsylvania  16802, USA}   
\author{J.~W.~Qiu}
\email{jqiu@bnl.gov}
\affiliation{Physics Department, Brookhaven National Laboratory,
  Upton New York 11973, USA}
\affiliation{C.N.~Yang Institute for Theoretical Physics, Stony Brook University,
  Stony Brook New York 11794, USA}
\author{T.~C.~Rogers}
\email{rogers@insti.physics.sunysb.edu}
\affiliation{Department of Physics and Astronomy, Vrije Universiteit Amsterdam, NL-1081 HV Amsterdam, The Netherlands}  
\affiliation{C.N.~Yang Institute for Theoretical Physics, Stony Brook University,
  Stony Brook NY 11794, USA}
\date{22 February 2012}

\begin{abstract}
  We extend the Collins-Soper-Sterman (CSS) formalism to apply it to
  the spin-dependence governed by the Sivers function.  We use it to
  give a correct numerical QCD evolution of existing fixed-scale fits
  of the Sivers function.  With the aid of approximations useful for
  the nonperturbative region, we present the results as
  parametrizations of a Gaussian form in transverse-momentum space,
  rather than in the Fourier conjugate transverse coordinate space
  normally used in the CSS formalism.  They are specifically valid at
  small transverse momentum.  Since evolution has been applied, our
  results can be used to make predictions for Drell-Yan and
  semi-inclusive deep inelastic scattering at energies different from
  those where the original fits were made.  Our evolved functions are
  of a form that they can be used in the same parton-model
  factorization formulas as used in the original fits, but now with a
  predicted scale dependence in the fit parameters. We also present a
  method by which our evolved functions can be corrected to allow for
  twist-3 contributions at large parton transverse momentum.
\end{abstract}
\maketitle

\section{Introduction}
\label{intro}
High energy collisions with transversely polarized hadrons
are ideal processes for extracting information about the 
structure of hadrons.  The nonperturbative functions that enter 
into the corresponding factorization 
formulas are sensitive to novel aspects of QCD dynamics such 
as
chiral symmetry breaking and the role of orbital angular momentum.
(See e.g.~\cite{Sivers:2011ci} for some interesting recent discussions.)
The Sivers function is an example which
has received considerable attention in recent years, and will 
be the focus of this article, although many of the results and techniques are extendable 
to other interesting transverse-momentum dependent (TMD) functions.
In loose terms, the Sivers function describes the transverse-momentum
distribution of (unpolarized)
partons inside a transversely polarized hadron
(usually a proton).  
In 
semi-inclusive 
cross sections with a single transversely 
polarized target hadron, it leads to a characteristic
$\sin( \phi - \phi_h)$  azimuthal modulation
($\phi$ and $\phi_h$ being the azimuthal angles of the transverse spin and the 
produced hadron, respectively).  
It is one of a 
collection of TMD
parton distribution functions (PDFs) and fragmentation functions (FFs)
that are actively being studied for the insight
they can provide about hadron structure and the unique 
opportunities they
provide
for comparing theoretical descriptions 
with experimental results~\cite{Boer:2011fh,rhicexps1,rhicexps2,Lu:2011cw,Kang:2011hk}.

The Sivers effect was originally proposed more than two 
decades ago in Ref.~\cite{Sivers:1989cc} as a mechanism for generating transverse single spin
asymmetries (SSAs) in hadron-hadron collisions.
Shortly afterward, it was argued in 
Ref.~\cite{Collins:1992kk} on the basis of time-reversal (actually $TP$)
invariance that the Sivers function vanishes.  This result, if true,
implies that the corresponding SSA in semi-inclusive deep inelastic
scattering (SIDIS) is power-suppressed (i.e., it is of ``higher
twist''), leaving only the spin-dependent effects due to the Collins
function in fragmentation.  Thus, a contradiction arose when spectator
model calculations \cite{BHS1} gave an explicit nonvanishing 
leading-twist SSA in SIDIS with the azimuthal dependence associated with the
Sivers function.  The situation was clarified in
Ref.~\cite{Collins:2002kn}, where it was shown that the proof of
vanishing of the Sivers function was incorrect in QCD, because it
ignored the Wilson lines needed in the definitions of parton
densities.  Instead, the true consequence of $TP$ invariance of QCD is
that the Sivers function reverses sign between SIDIS and Drell-Yan (DY)
processes.  This is because future-pointing Wilson lines are needed in
TMD functions like the Sivers function when used for SIDIS, but
past-pointing Wilson lines are needed for the Drell-Yan process.  At
the level of the actual cross section, the sign-reversal for the
Drell-Yan process was verified in model calculations in
Ref.~\cite{BHS2}.

Certain other polarization or azimuthally dependent functions, such as
the Boer-Mulders and the pretzelosity distributions
\cite{Boer:1998nt,Avakian:2008dz}, also share this 
``T-odd'' property of reversal of sign between SIDIS and Drell-Yan.
Over the past decade, there has developed much work in the extraction,
study, and formal theoretical description of these functions.

However, phenomenological fits of the Sivers function (and of related
functions) have so far \cite{Collins:2005ie,Anselmino:2008sga} used
only the simplest parton-model factorization formulas where the TMD
parton densities and fragmentation functions do not evolve with the
scale of the process, 
or use incorrect evolution formalisms.  
This is inadequate when they are to be applied
to experiments at widely different energies.  There is a good QCD
formalism for applying TMD functions in a factorization framework, due
to Collins, Soper and Sterman (CSS)~\cite{CS1,CSS1}.  The CSS
formalism gives a correct treatment of the region of low transverse
momentum, which is where the Sivers function analysis is used.
However it has not been fully systematized for the case of the Sivers
function and other azimuthally-dependent functions, except in the work
of Boer \cite{Boer:2001he,Boer:2008fr} 
and Idilbi et al.~\cite{Idilbi:2004vb}, 
on which we comment below.

In this paper, we give a complete extension of the CSS method to
processes that need the Sivers function, using the methods recently given
in Ref.~\cite{collins}.  It is straightforward to extend our results
to the other azimuthally dependent PDFs and FFs (e.g., the Collins
function and the Boer-Mulders function).  We apply the formalism to
give numerical results for the Sivers function evolved from existing
fits.  
The only extra nonperturbative information needed for the
evolution is 
universal and is 
obtained from existing fits to the unpolarized Drell-Yan
process.  This extends the results given by two of us in
Ref.~\cite{Aybat:2011zv} for the unpolarized case.
Reference~\cite{Anselmino:2008sga} attempts to include some effects of evolution 
by simply including the evolution from collinear factorization, but
this is
incorrect for TMD-factorization.  It is also stated (Ref.~\cite{Anselmino:2008sga}, for example) 
that the true scale-evolution 
of the Sivers function is unknown.  One purpose of this
article is to demonstrate that this is no longer true.  

With the aid of an approximation useful for the nonperturbative region, 
we present the results as Gaussian transverse-momentum distributions
with scale-dependent parameters.  They are therefore as easy to use in
simple parton-model-style calculations as the original fixed-scale
fits \cite{Collins:2005ie,Anselmino:2008sga}.  As the scale increases,
the distributions broaden substantially in transverse momentum, and
get diluted in size.  It will be necessary to include perturbative
twist-3 corrections to get more accurate values at the larger values
of transverse momentum, and we present a scheme for how this should be
done.  

Boer \cite{Boer:2001he,Boer:2008fr} has applied the CSS method to
processes involving the Collins function.  
Idilbi et al.~\cite{Idilbi:2004vb} have applied the CSS method to 
their definitions of various TMD distributions~\cite{Ji:2004xq,Ji:2004wu} including the Sivers function.
Our treatment is
substantially improved, to include a correct treatment of the
nonperturbative region in CSS evolution applied to T-odd functions,
to use a more modern version of the CSS formalism, to apply it to the
Sivers function, and to obtain convenient numerical results for the
Sivers function.

Although it has recently become common for the word ``resummation'' to
be used to indicate any CSS-like treatment, in our work we will
maintain a firm distinction between resummation methodology and
TMD-factorization.
The term ``resummation'' is often used to indicate that one starts
with conventional collinear factorization and resums logarithms of
$q_T/Q$, which can in fact be done with the CSS methodology.  The
problem with this approach is that it is only valid when the
underlying collinear factorization formula is valid, i.e., for the
region where the transverse momentum $q_T$ is both much less than the
hard scale and much greater than hadronic binding energies $\sim \Lambda_{\rm
  QCD}$.  (See, in particular, the recent work of
Ref.~\cite{Kang:2011mr}.)  But to extend the calculations to
transverse momenta comparable to $\Lambda_{\rm QCD}$ and to zero transverse
momentum requires a complete TMD-factorization formalism, which we use
here.  This is particularly important because many SIDIS experiments
such as HERMES and JLab are performed at kinematical scales where
transverse momenta of order $\Lambda_{\rm QCD}$ are certainly important,
and $Q$ is not so large.  

A number of difficulties are caused by the use of a pure resummation
formalism rather than TMD factorization as the basis of calculations.
For the present paper, one of the most significant is that a
leading-power resummation formalism does not give the effects
associated with the Sivers function (and also those associated with
the Boer-Mulders \cite{Boer:1997nt} function).  But, provided that
spin effects are treated correctly, the presence of these functions is
automatic in TMD factorization, at leading power.

\section{Setup and Definitions}
\label{sec:tmdfactorization}
In this section we give the factorization formula for SIDIS: $e+P(S)\to
e+h+X$, and present the definitions of the TMD functions.  We let $P$
and $S$ be the momentum and spin vector of the hadron target, and we
let $h$ label the detected hadron, of momentum $p_h$.  With a single
exchanged photon of momentum $q$, independent kinematic variables are:
$Q=\sqrt{-q^2}$, $x=Q^2/2p\cdot q$, $z=P\cdot p_h/P\cdot q$, and the virtual
photon's transverse momentum ${\bf q}_{\rm T}$ (in a hadron frame
where the measured hadrons have zero transverse momentum).

\begin{widetext}
The TMD-factorization formula in the form derived by Collins
\cite{collins} is:
\begin{align}
\label{eq:parton2}
   W^{\mu \nu} ={}& \sum_f \left| \mathcal{H}_{f}(Q;\mu)^2 \right|^{\mu \nu} \, 
      \int d^2 {\bf k}_{1\rm T} \, d^2 {\bf k}_{2\rm T} \, 
      F_{f/P^{\uparrow}}(x,{\bf k}_{1\rm T},S;\mu;\zeta_F) \, 
      D_{h/f}(z,z {\bf k}_{2\rm T};\mu;\zeta_D) \,
      \delta^{(2)}({\bf k}_{1\rm T} + {\bf q}_{\rm T} - {\bf k}_{2\rm T})
\nonumber\\&
 + Y(Q,\trans{q}) + \mathcal{O}((\Lambda / Q)^a).
\end{align} 
\end{widetext}
Here $F_{f/P^{\uparrow}}(x,{\bf k}_{1T},S)$ is the TMD PDF for an
unpolarized quark of flavor $f$ in a proton
of polarization $S$, and $D_{h/f}(z,z {\bf k}_{2T})$ is the
unpolarized fragmentation function.  These factors contain
nonkinematic parameters, $\mu$, $\zeta_F$, and $\zeta_D$, whose definitions
are given below.  The hard-scattering factor $|\mathcal{H}^2|^{\mu\nu}$
is computed, with appropriate subtractions, from massless parton
scattering in a photon frame where the photon and partons have zero
transverse momentum --- see \cite[page 527]{collins} for its
definition.  The first line of the factorization formula is valid at
low transverse momentum, and the $Y$ term provides a correction for
large transverse momentum in a form like that for ordinary collinear
factorization.
Although we will focus on SIDIS for this paper, the same 
general treatment applies also to DY scattering, up to the 
change in direction of the Wilson line in the definition of 
the TMD PDF.  
Note that the TMD-factorization piece, the first term in Eq.~\eqref{eq:parton2},
is formulated specifically to deal with the small $k_T$ behavior ($k_T \to 0$), while allowing 
for systematic corrections to the behavior as $k_T$ grows larger than $\Lambda_{\rm QCD}$.

The above formula is exactly like the parton-model formula for the
same cross section except for the scale dependence of the PDF and
fragmentation function and except for higher-order corrections in the
hard scattering 
and $Y$-term.  
It differs from the older CSS formula by no longer
needing an explicit soft factor.
The factorization formula~\eqref{eq:parton2} is written for the case
that the partons at the hard scattering are unpolarized.  Parton
polarization effects can be allowed for simply by inserting spin
matrices for the incoming and outgoing partons of the hard scattering.
This gives other terms, e.g., those with the Collins function in
fragmentation, with their characteristic angular distributions in the
cross section.  
It was recently suggested in Ref.~\cite{Boer:2011xd} that it would be useful to analyze data for cross sections
in transverse coordinate space $b_T$ by taking various weighted
integrals with Bessel functions.  In that case, the $b_T$ version of Eq.~\eqref{eq:parton2} is needed.

The parameter $\mu$ is a conventional renormalization scale, which we
will choose to be in the \MSbar{} scheme.  It should be chosen to be
of order $Q$ so that the hard scattering has no large logarithms.  The
parameters $\zeta_F$ and $\zeta_D$ are related to the need to regulate
rapidity divergences in the definitions of the TMDs.  They are defined
with the aid of an auxiliary rapidity parameter $y_s$, which has the
function of separating forward and backward rapidity gluons.  We use a
hadron frame (in which the hadrons have zero transverse momentum),
oriented so that $e^{y_P} \gg e^{y_{p_h}}$, and we let $M_P$ and $M_h$
be the masses of these hadrons.  Then $\zeta_F$ and $\zeta_D$ are defined by
\begin{equation}
\label{eq:zetaF}
\zeta_F = M_P^2 x^2 e^{2(y_P - y_s)}
\end{equation}
and 
\begin{equation}
\label{eq:zetaD}
\zeta_D = (M_h^2 /  z^2) e^{2(y_s - y_h)}.
\end{equation}
They obey $\sqrt{\zeta_F \zeta_D} = Q^2$ up to power-suppressed corrections,
and have been normalized to correspond to CSS's definitions.

The definitions of gauge-invariant TMD functions are equipped with
Wilson lines.  A Wilson line (or gauge link) from a point $x$ to $\infty$
along the direction of a four-vector $n$ is defined as
\begin{equation}
\label{eq:wildef}
W(\infty ,x;n) = P \exp \left[- ig_0 \int_0^\infty d s \; n \cdot A_0^a (x + s n) t^a \right].
\end{equation}
Here, bare field operators and bare couplings are used and $P$ is a
path-ordering operation.  The generator for the gauge group in the
fundamental representation, with color index $a$, is denoted by $t^a$.

To define the parton densities, we use two lightlike directions that
characterize the extreme forward and backward directions:
\begin{equation}
\label{eq:LLdir}
u_{\rm A} = (1,0,{\bf 0}_T), \qquad u_{\rm B} = (0,1,{\bf 0}_T).
\end{equation} 
These correspond to the directions of $P$ and $p_h$.  
Our coordinates for a 4-vector $v$ are defined by 
\begin{equation}
v=(v^+,v^-,v_T)
\end{equation} 
where, 
\begin{equation}
v^{\pm}=(v^0\pm v^z)/\sqrt{2}.
\end{equation}

Now the most
obvious definitions of PDFs use light-like Wilson lines, which give
rise to rapidity divergences \cite{Collins:2008ht}.  Regulating the
divergences can be done 
by using non-light-like Wilson lines.  So we define vectors $n_{\rm
  A}(y_{\rm A})$ and $n_{\rm B}(y_{\rm B})$ with finite rapidities
$y_A$ and $y_B$:
\begin{equation}
\label{eq:nLLdir}
    n_{\rm A} = (1,-e^{-2 y_{\rm A}},{\bf 0}_T),
\qquad 
    n_{\rm B} = (-e^{2 y_{\rm B}},1,{\bf 0}_T).
\end{equation}

The actual TMD PDF in Eq.~\eqref{eq:parton2} is defined as a limit of
an unsubtracted TMD multiplied by certain unsubtracted soft functions.
These are
first defined in transverse coordinate space and then the final result
will be Fourier transformed to transverse-momentum space.  The
unsubtracted TMD PDF is
\begin{widetext}
\begin{multline}
\label{eq:def1}
\tmdpdf_{f/P^{\uparrow}}^{\rm unsub}(x,\trans{b},S;\mu;y_P - y_B) \\
= {\rm Tr}_{C} {\rm Tr}_D \int \frac{d w^{-}}{2 \pi} e^{-i x P^+ w^-} 
   \langle P,S | \bar{\psi}_f(w/2) W(w/2,\infty,n_B(y_B))^\dagger 
  \frac{\gamma^+}{2} W(-w/2,\infty,n_B(y_B)) \psi_f(-w/2) | P ,S \rangle_{c}
\end{multline}
where $w^\mu = (0^+, w^-, {\bf b}_T)$, and we notate the functions with
a tilde to indicate the use  
of transverse coordinate space. The subscript $c$ indicates that only connected diagrams
are included, and ${\rm Tr}_{C}$ and ${\rm Tr}_D$ represent color and
Dirac traces respectively.  The unsubtracted soft function is
\begin{equation}
\label{eq:soft}
  \tilde{S}_{(0)}(\trans{b};y_A,y_B) = 
  \frac{1}{N_c} \langle 0 |W(\trans{b}/2,\infty;n_B)^{\dagger}_{ca} \, 
                    W(\trans{b}/2,\infty;n_A)_{ad} 
                    W(-\trans{b}/2,\infty;n_B)_{bc}
                    W(-\trans{b}/2,\infty;n_A)^{\dagger}_{db} | 0 \rangle.
\end{equation}
In both of these functions, there should be inserted transverse gauge
links at infinity.  However, their effects cancel in the subtracted TMD
PDF, when Feynman gauge is used, so we have not indicated the extra
gauge links explicitly.

The full definition of the TMD PDF from~\cite{collins} is
\begin{equation}
\label{eq:def2}
\tmdpdf_{f/P^\uparrow}(x,\trans{b},S;\mu,\zeta_F) = \tmdpdf^{\rm unsub}_{f/P^\uparrow}(x,\trans{b},S;\mu;y_P - (-\infty)) 
\sqrt{\frac{\tilde{S}_{(0)}(\trans{b};+\infty,y_s)}{\tilde{S}_{(0)}(\trans{b};+\infty,-\infty) \tilde{S}_{(0)}(\trans{b};y_s,-\infty)}} Z_F \, Z_2.
\end{equation}
This involves limits: infinite rapidity on the Wilson lines indicated,
infinite length for the Wilson lines, and then removal of the UV
regulator (dimensional regularization).  The factors $Z_F Z_2$ at the
end of Eq.~\eqref{eq:def2} are the field strength and TMD
renormalization factors respectively.  Notice that two of the soft
factors have one of their rapidity arguments equal to the finite
parameter $y_s$. 

An exactly 
analogous 
definition applies to the fragmentation function 
(see Ref.~\cite{collins} for the explicit definition).
In our notation, capital letters will denote unintegrated quantities and lower case letters will denote quantities integrated over 
transverse momentum.
Otherwise, we will stick as closely as possible to the Trento conventions~\cite{trento}.

The momentum-space TMD PDF is
\begin{equation}
\label{eq:def3}
F_{f/P^\uparrow}(x,\trans{k},S;\mu,\zeta_F) = 
\frac{1}{(2 \pi)^2} \int d^2 {\bf b}_T \, e^{i {\bf k}_T \cdot {\bf b}_T} \, \tmdpdf_{f/P^\uparrow}(x,\trans{b},S;\mu,\zeta_F).
\end{equation}
This has dependence on the azimuthal angle between ${\bf k}_T$ and the
transverse spin vector ${\bf S}_T$ of the target hadron. (We normalize
${\bf S}_T$ so that its maximum size is unity.)
The TMD PDF is decomposed as
usual into the unpolarized TMD PDF and a spin-dependent term:
\begin{equation}
\label{eq:polpdf}
F_{f/P^{\uparrow}}(x,k_T,S;\mu,\zeta_F) =  F_{f/P}(x,k_T;\mu,\zeta_F) 
- F^{\perp \,f}_{1T}(x,k_T;\mu,\zeta_F) \frac{\epsilon_{ij}  k_T^i S^j}{M_p},
\end{equation} 
with $F^{\perp \,f}_{1T}(x,k_T;\mu,\zeta_F)$ being the Sivers function.
\end{widetext}

\section{Evolution of the Sivers Function}
\label{sec:evol}
In this section we generalize CSS evolution from the
unpolarized TMDs to the Sivers function.  Similar methods apply to the
other TMDs with azimuthal dependence.

The general CSS formalism works equally well for these functions
\cite{collins}.  But it involves Fourier transformations in two
transverse dimensions, and for practical use it is convenient to
perform the azimuthal integrals analytically and 
to write the transforms
in terms of integrals over the sizes of the transverse variables.
The treatment of the azimuthal integrals provided in Sec.~\ref{sec:coord} closely parallels previous treatments in 
Refs.~\cite{Ji:2004xq,Idilbi:2004vb} and recently in~\cite{Boer:2011xd}. 

\subsection{Coordinate Space Representation of Azimuthal Dependence}
\label{sec:coord}
To analyze the evolution of the last term in Eq.~\eqref{eq:polpdf} we
extract the azimuth-dependent part by defining
\begin{equation}
\label{eq:azimdep}
\phi^i_{f/P}(x,\trans{k};\mu,\zeta_F)  
\equiv \frac{k_T^i}{M_p} \, F^{\perp \,f}_{1T}(x,k_T;\mu,\zeta_F),
\end{equation}
in terms of which the complete Sivers term is
\begin{equation}
\label{eq:azimdep2}
   F^{\perp \,f}_{1T}(x,k_T;\mu,\zeta_F) \frac{\epsilon_{ij}  k_T^i S_T^j}{M_p}
   =
   \phi^i_{f/P}(x,\trans{k};\mu,\zeta_F) \epsilon_{ij} S_T^j.
\end{equation}

The Fourier transform of the Sivers function is
\allowdisplaybreaks
\begin{align}
\label{eq:azimdep3}
   \tilde{F}^{\perp \,f}_{1T}(x,b_T;\mu,\zeta_F) 
   &= \int d^2 \trans k \,e^{-i \trans k\cdot \trans b} \, F^{\perp \,f}_{1T}(x,k_T;\mu,\zeta_F)
\nonumber\\
   &\hspace*{-2cm}
   =  2 \pi \int_0^\infty \, d k_T \,  k_T J_0(k_T b_T) F^{\perp \,f}_{1T}(x,k_T;\mu,\zeta_F),
\end{align}
and the Fourier transform of $\phi^i_{f/P}(x,\trans{k};\mu,\zeta_F)$ is
\begin{align}
   \tilde{\phi}^i_{f/P}(x,\trans{b};\mu,\zeta_F) 
   &= \int d^2\trans{k}\,e^{-i\trans{k}\cdot\trans{b}}\phi^i_{f/P}(x,\trans{k};\mu,\zeta_F)
\nonumber\\
   &\hspace*{-2.5cm}
   = \int d^2\trans{k}\,e^{-i\trans{k}\cdot\trans{b}}\, \frac{k_T^i}{M_p}
     \, F^{\perp \,f}_{1T}(x,k_T;\mu,\zeta_F)
\nonumber\\
   &\hspace*{-2.5cm}
   = \frac{1}{M_P}\int d^2\trans{k}\,
     \frac{i\partial}{\partial b_{Ti}}\,e^{-i\trans{k}\cdot\trans{b}}\, 
    F^{\perp \,f}_{1T}(x,k_T;\mu,\zeta_F)\,.
\end{align}
Using Eq.~\eqref{eq:azimdep3} gives
\begin{equation}
\label{eq:bspace_phi}
\tilde{\phi}^i_{f/P}(x,\trans{b};\mu,\zeta_F) = i\,\frac{1}{M_P}\,\frac{b_T^i}{b_T}\,\tilde{F}^{\prime \, \perp \,f}_{1T}(x,b_T;\mu,\zeta_F)\,,
\end{equation}
where we have denoted the derivative of $\tilde{F}^{\perp \,f}_{1T}$ with
respect to the length of ${\bf b}_T$ by
\begin{equation}
\label{eq:azimdep4}
\tilde{F}^{\prime \, \perp \,f}_{1T}(x,b_T;\mu,\zeta_F) \equiv \frac{\partial \tilde{F}^{\perp \,f}_{1T}(x,b_T;\mu,\zeta_F)}{\partial b_T}.
\end{equation}

As we will see shortly, it is this derivative $\tilde{F}^{'}$ and not
the function $\tilde{F}$ itself that gets used in the evolution
equations and in the formula for the Sivers term in the actual
transverse-momentum dependence in Eq.~(\ref{eq:polpdf}). 

Taking an inverse Fourier transform of Eq.~\eqref{eq:bspace_phi} 
allows $\phi^i_{f/P}(x,\trans{k};\mu,\zeta_F)$ to be rewritten in terms of Eq.~\eqref{eq:azimdep4}:
\begin{align}
   \phi^i_{f/P}(x,\trans{k};\mu,\zeta_F)
\nonumber\\ 
 &\hspace*{-2cm}
   = \frac{1}{(2\pi)^2}\int d^2\trans{b}\,e^{i\trans{k}\cdot\trans{b}}\,
     \tilde{\phi}^i_{f/P}(x,\trans{b};\mu,\zeta_F) 
\nonumber\\ 
 &\hspace*{-2cm}
    =  \frac{i}{(2\pi)^2M_P}\int d^2\trans{b}\,e^{i\trans{k}\cdot\trans{b}}\,
      \frac{b_T^i}{b_T}\, \tilde{F}^{\prime \, \perp \,f}_{1T}(x,b_T;\mu,\zeta_F)\,.
\end{align}
To further simplify this expression, and without loss of generality, we use a frame where $\trans{k}$ 
is in the $x$ direction so that $\frac{k_T^i}{k_T}=(1,0)$ and $\frac{b_T^i}{b_T}=(\cos\theta,\sin\theta)$. 
Then,
\begin{widetext}
\begin{eqnarray}
\label{eq:azimdep5}
  \phi^i_{f/P}(x,\trans{k};\mu,\zeta_F) 
  &=&  \frac{i}{(2\pi)^2M_P}\int_0^{\infty} db_T\,b_T\, 
  \tilde{F}^{\prime \, \perp \,f}_{1T}(x,b_T;\mu,\zeta_F)\int_{-\pi}^{\pi}d\theta\,e^{ik_Tb_T\cos\theta}(\cos\theta,\sin\theta)\,
\nonumber\\
  &=&\frac{1}{(2\pi)^2M_P}\int_0^{\infty} db_T\,b_T\, 
  \tilde{F}^{\prime \, \perp \,f}_{1T}(x,b_T;\mu,\zeta_F)\frac{\partial}{\partial (k_Tb_T)}\int_{-\pi}^{\pi}d\theta\,e^{ik_Tb_T\cos\theta}(1,0)\,
\nonumber\\
  &=&\frac{k_T^i}{2\pi M_P k_T}\int_0^{\infty} db_T\,b_T\, 
  \tilde{F}^{\prime \, \perp \,f}_{1T}(x,b_T;\mu,\zeta_F)\frac{\partial}{\partial (k_Tb_T)}J_0(k_Tb_T)\,
\nonumber\\
  &=&  \frac{-k_T^i}{2 \pi M_p k_T} 
  \int_0^\infty d b_T \, b_T J_1(k_T b_T) \tilde{F}^{\prime \, \perp \,f}_{1T}(x,b_T;\mu,\zeta_F)\,.
\end{eqnarray}
Then the complete Sivers term in Eq.~(\ref{eq:polpdf}) is
\begin{equation}
\label{eq:azimdep6}
\phi^i_{f/P}(x,\trans{k};\mu,\zeta_F) \epsilon_{ij} S_T^j
= \frac{-k_T^i \epsilon_{ij} S_T^j}{2 \pi M_p k_T} \int_0^\infty d b_T \, b_T J_1(k_T b_T) \tilde{F}^{\prime \, \perp \,f}_{1T}(x,b_T;\mu,\zeta_F).
\end{equation}
So, from Eq.~\eqref{eq:azimdep2} we express the momentum-space Sivers
function in terms of $\tilde{F}^{'}$:
\begin{equation}
\label{eq:azimdep7}
F^{\perp \,f}_{1T}(x,k_T;\mu,\zeta_F) = 
\frac{-1}{2 \pi k_T} \int_0^\infty d b_T \, b_T J_1(k_T b_T) \tilde{F}^{\prime \, \perp \,f}_{1T}(x,b_T;\mu,\zeta_F).
\end{equation}
whose inverse transform is
\begin{equation}
\label{eq:azimdep8}
\tilde{F}^{\prime \, \perp \,f}_{1T}(x,b_T;\mu,\zeta_F) = 
-2 \pi \int_0^\infty d k_T \, k_T^2 J_1(k_T b_T) F^{\perp \,f}_{1T}(x,k_T;\mu,\zeta_F).
\end{equation}
Notice that the originally defined $\tilde{F}^{\perp \,f}_{1T}$ from
Eq.~(\ref{eq:azimdep3}) no longer appears.  The $b_T$-dependent function $\tilde{F}^{\prime \, \perp \,f}_{1T}(x,b_T;\mu,\zeta_F)$  is 
closely analogous to the quantity $\tilde{f}^{\perp(1)}_{1T}$ that
appears in Eqs.\ (16) and (20) of Ref.~\cite{Boer:2011xd}, and to $\partial^i_b
q_T$ in Eq.~(40) of Ref.~\cite{Idilbi:2004vb}, though  
the basic definition for the $b_T$-space TMD PDF in
Eq.~\eqref{eq:def2} is significantly different. 
\end{widetext}

\subsection{The Evolution Equations}
\label{sec:eveqs}
The set of evolution equations comprises the Collins-Soper (CS) equation
which gives evolution with respect to $\zeta_F$, and the RG equations
which give evolution with respect to $\mu$.  The CS equation for the
TMD function defined in Eq.~\eqref{eq:def2} is \cite{collins}
\begin{multline}
\label{eq:origevol}
\frac{\partial \tmdpdf_{f/P^\uparrow}(x,\trans{b},S;\mu,\zeta_F)}{\partial \ln \sqrt{\zeta_F} } 
= 
\\
\tilde{K}(b_T;\mu) \tmdpdf_{f/P^\uparrow}(x,\trans{b},S;\mu,\zeta_F),
\end{multline}
where
\begin{equation}
\label{eq:KPDF}
\tilde{K}(b_T;\mu) = \frac{1}{2} \frac{\partial}{\partial y_s} \ln \left( \frac{\tilde{S}(b_T;y_s,-\infty)}{\tilde{S}(b_T;+\infty,y_s)} \right).
\end{equation}
The RG equations are
\begin{equation}
\label{eq:RGKPDF}
\frac{d \tilde{K}(b_T;\mu)}{d \ln \mu} = - \gamma_K(g(\mu))
\end{equation}
and
\begin{multline}
\label{eq:RGPDF}
\frac{d \tmdpdf_{f/P^\uparrow}(x,\trans{b},S;\mu,\zeta_F)}{d \ln \mu} \\
= \gamma_F(g(\mu);\zeta_F /\mu^2) \tmdpdf_{f/P^\uparrow}(x,\trans{b},S;\mu,\zeta_F).
\end{multline}
Similar equations apply to the fragmentation function.

It follows that the $\zeta_F$ dependence of $\gamma_F$ is determined:
\begin{equation}
\label{eq:gam.F.evol}
  \frac{ \partial \gamma_F(g(\mu);\zeta_F /\mu^2) }{ \partial \ln \sqrt{\zeta_F} } 
= 
  -\gamma_K(g(\mu)) ,
\end{equation}
so that
\begin{equation}
\label{eq:gam.F.zeta.dep}
  \gamma_F(g(\mu);\zeta_F /\mu^2)
= 
  \gamma_F(g(\mu);1)
  - \frac12 \gamma_K(g(\mu)) \ln \frac{\zeta_F}{\mu^2}.
\end{equation}

These equations were used in Ref.~\cite{Aybat:2011zv} to calculate the
evolution of the unpolarized TMDs.  For the spin-dependent case, the
Fourier transform of the second term in Eq.~\eqref{eq:polpdf} obeys the
same evolution equations, i.e., the equations apply to
\begin{multline}
\label{eq:evolpart1}
\int d^2 \trans k \,e^{-i \trans k\cdot \trans b} \, F^{\perp \,f}_{1T}(x,k_T;\mu,\zeta_F) \frac{\epsilon_{ij} k_T^i S_T^j}{M_p}
\\
= \tilde{\phi}^i_{f/P}(x,\trans{b};\mu,\zeta_F) \epsilon_{ij} S_T^j.
\end{multline} 
The CS equation for the spin-dependent part is therefore
\begin{multline}
\label{eq:spinpart2}
\frac{\partial \tilde{\phi}^i_{f/P}(x,\trans{b};\mu,\zeta_F) \epsilon_{ij} S_T^j}{\partial \ln \sqrt{\zeta_F} } \\
= \tilde{K}(b_T;\mu) \tilde{\phi}^i_{f/P}(x,\trans{b};\mu,\zeta_F) \epsilon_{ij} S_T^j\,.
\end{multline}
Hence, Eq.~\eqref{eq:bspace_phi} shows that the CS equation for
$\tilde{F}^{\prime \, \perp \,f}_{1T}(x,b_T;\mu,\zeta_F)$ is the same
as for the unpolarized TMD PDF:
\begin{equation}
\label{eq:spinpart4}
\frac{\partial  \ln \tilde{F}^{\prime \, \perp \,f}_{1T}(x,b_T;\mu,\zeta_F) }{\partial \ln \sqrt{\zeta_F} } 
= \tilde{K}(b_T;\mu).
\end{equation}
Similarly, its RG equation is like Eq.~\eqref{eq:RGPDF}:
\begin{multline}
\label{eq:spinpart5}
\frac{d  \tilde{F}^{\prime \, \perp \,f}_{1T}(x,b_T;\mu,\zeta_F) }{d \ln \mu} 
\\ = \gamma_F(g(\mu);\zeta_F /\mu^2) \tilde{F}^{\prime \, \perp \,f}_{1T}(x,b_T;\mu,\zeta_F).
\end{multline}

Note that in Eqs.~(\ref{eq:spinpart4}) and (\ref{eq:spinpart5}) the same CS
kernel $\tilde{K}(b_T;\mu)$ and anomalous dimension $\gamma_F(g(\mu);\zeta_F
/\mu^2)$ appear as in the unpolarized case.  This is because
$\tilde{K}$ and $\gamma_F$ are properties of the operator defining the
parton density, and this operator is the same for the ordinary
unpolarized TMD PDF as for the Sivers function; both concern the
number density of quarks in a hadron, with no polarization restriction
on the quark.

It is important to emphasize that the evolution equations
(\ref{eq:origevol},~\ref{eq:RGKPDF},~\ref{eq:RGPDF}) are set up to be
exactly correct for all $b_T$, and for all $k_T$.  This includes the
region where $b_T \to \infty$ (and hence $k_T \to 0$).  Indeed, the first
term on the right side of Eq.~\eqref{eq:parton2} (the
TMD-factorization term) is designed to give an accurate pQCD treatment
when $k_T \ll Q$, independently of the relative sizes of $k_T$ and
$\Lambda_{\rm QCD}$.

\subsection{Power laws for $k_T$ and $b_T$ dependence}
\label{sec:powerlaws}

As a guide to the qualitative behavior of the Sivers function, we
summarize in this section the power laws for its dependence on
transverse momentum and transverse position as obtained from simple
model calculations.
(For a detailed treatment of the power law behavior of other TMDs, 
see Ref.~\cite{Bacchetta:2008xw} and 
also recent discussions in Ref.~\cite{Boer:2011xd}.)
In purely perturbative higher-order calculations,
these get modified by logarithms, while use of a correct solution of
the evolution equations can significantly modify the power laws
\cite{Collins:1981va}.  Nevertheless, the power laws from elementary
perturbative calculations form a useful standard of comparison.

First, we characterize the power law for an ordinary unpolarized TMD
PDF by
\begin{equation}
  \label{eq:unpol.kt.power}
  F(x,k_T) \sim \frac{1}{k_T^2+M^2}.
\end{equation}
At large $k_T$, the falloff $1/k_T^2$ is the simple
dimensional-analysis power, appropriate to a theory with a
dimensionless coupling.  The increase at low $k_T$ is tamed by an
infra-red cutoff $M$, which in QCD is nonperturbative.  In
$b_T$ space, the large-$k_T$ behavior Fourier transforms to
\begin{equation}
  \label{eq:unpol.bt.power}
  \tilde{F}(x,b_T) \sim \mbox{constant} \times \mbox{logarithms}
\quad (\mbox{as $b_T\to 0$}).
\end{equation}
At large $b_T$, the falloff of $\tilde{F}$ should be at least rapid
enough that the integral over all $b_T$ is convergent, to give a finite
value for $F(x,k_T)$ at $k_T=0$.  Normally an exponential or Gaussian
falloff is assumed (which is controlled by nonperturbative effects in
QCD).

As for the Sivers function, its contribution to the quark density ,
$F^{\perp \,f}_{1T}(x,k_T) {\epsilon_{ij} k_T^i S^j}/{M_p}$, has a
kinematic zero at $k_T=0$.  In addition, it is a chirality-violating
quantity, and at large $k_T$, this
requires a suppression by a factor of mass divided by $k_T$ relative
to the unpolarized density.  So we characterize the result by
\begin{equation}
  \label{eq:Sivers.kt.power1}
  F^{\perp \,f}_{1T}(x,k_T) \frac{\epsilon_{ij}  k_T^i S^j}{M_p}
  \sim \frac{k_TM}{ (k_T^2+M^2)^2 }.
\end{equation}
For the Sivers function itself, we therefore have
\begin{equation}
  \label{eq:Sivers.kt.power2}
  F^{\perp \,f}_{1T}(x,k_T)
  \sim \frac{M^2}{ (k_T^2+M^2)^2 }.
\end{equation}
This falloff is characterized as ``twist-3.''  In $b_T$ space, the
behavior of the Fourier transform of (\ref{eq:Sivers.kt.power2}) at
small $b_T$ is 
\begin{align}
  \label{eq:Sivers.bt.power1}
  \tilde{F}^{\perp \,f}_{1T}(x,b_T)
  \sim \mbox{constant} + b_T^2 \times \mbox{logarithms}.  
\end{align} 
However, as we saw, it is the \emph{derivative} of this quantity with
respect to $b_T$ that is actually used, for which the behavior is
linear:
\begin{align}
  \label{eq:Sivers.bt.power2}
  \tilde{F}^{\prime \, \perp \,f}_{1T}(x,b_T)
  \sim b_T \times \mbox{logarithms}.  
\end{align}
Although the actual equations for evolution are the same for the
Sivers function as for the standard unpolarized TMD PDF, there
are substantial differences in the way in which the evolution is
reflected in the numerical values of these functions in
transverse-momentum space.  Because $\tilde{F}_{1T}^{'\perp}$
is approximately linear in $b_T$ at small $b_T$ and because
the $J_1$ Bessel function instead of $J_0$ appears in Eq.\ (21),
the Fourier transform for the Sivers function is sensitive to
larger $b_T$ values than the transform for the unpolarized TMD.
This also implies that the evolution of Sivers is subject to
more uncertainty from the nonperturbative large-$b_T$ region than
that of the unpolarized TMD.

\subsection{Small-$b_T$ expansion}

For the unpolarized TMD PDF, an expansion for small $b_T$ can be made
in terms of the integrated PDFs.  After Fourier transformation, this
gives both the large-$k_T$ behavior, and the normalization of the
integral over the whole small $k_T$ region.  

The same idea continues to apply when we include the dependence of the
TMD density on the target polarization.  We can write
\begin{equation}
\label{eq:small.b}
  \tilde{F}(x,{\bf b}_T,S)
  = \sum_j \mbox{coefficient}_j \otimes \langle P,S| \mbox{operator}_j |P,S\rangle,
\end{equation}
where the coefficients and operators are unaltered since they are
properties of the TMD number-density operator.  But the twist-2
operator on the right-hand side of \eqref{eq:small.b} is the ordinary
number-density operator used to define an 
integrated PDF, and its matrix element is independent of
transverse spin.  Thus the twist-2 operator, corresponding to a
$1/k_T^2$ fall off at large $k_T$, provides no contribution to the
Sivers function in Eq.~(\ref{eq:small.b}).  The leading large-$k_T$
behavior of the Sivers function is the $1/k_T^3$ term associated with
the twist-3 operators, the same operators that are used in the
Qiu-Sterman formalism \cite{Qiu:1998ia}.

\section{Obtaining evolved Sivers functions}

In this section, we discuss the steps for obtaining the evolved Sivers 
function using already existing fits to the nonperturbative parts.

\subsection{Solution in terms of fixed-scale Sivers function}

Previous fits \cite{Collins:2005ie,Anselmino:2008sga} of the Sivers
function used the parton-model formula for the hadronic tensor.  We
now show how these can be converted to use the correct QCD formula.  

The parton-model version of TMD factorization amounts to applying the
following approximations to the true QCD formula (\ref{eq:parton2}):
\begin{enumerate}
\item Replace the hard scattering by its lowest order.
\item Neglect the $Y$ term.
\item Omit the evolution of the TMD PDFs.
\end{enumerate}
If the renormalization scale $\mu$ is taken of order $Q$, higher-order
corrections to the hard scattering are purely perturbative.  One of
the simplifications for TMD factorization is that these are just an
overall factor, dependent on $Q$ only through the running coupling
$\alpha_S(Q)$.  This factor is the same, independent of the hadron and the
quark polarization, so it does not affect the ratio of the Sivers
function to the ordinary TMD PDF.

The $Y$ term only affects large transverse momentum (of order $Q$),
whereas the data is dominantly at transverse momenta in the
nonperturbative region.  So the neglect of $Y$ should be an
adequate approximation with present data, and is easily corrected in
the future, with the aid of fits for the Qiu-Sterman twist-3 function.

For a fixed value of $Q$, the TMD functions can be given fixed values
of $\mu$ and $\zeta_F$, $\mu=Q$ and $\zeta_F=Q^2$, and the QCD factorization
formula is the same as the parton-model formula, up to an overall
$K$-factor.  This legitimizes the fixed-scale fits.  But as can be
seen from Fig.\ \ref{fig:siversevolved} below,
evolution gives substantial changes in the TMD PDFs needed at higher
$Q$.  These are easily obtained, in their transverse-coordinate-space
form, in terms of the parton-model fits at a fixed scale.  We derive
the necessary result starting from Eqs.\ (\ref{eq:spinpart4}),
(\ref{eq:spinpart5}), and (\ref{eq:gam.F.zeta.dep}).

In these equations, the anomalous dimensions $\gamma_F$ and $\gamma_K$ are
perturbatively calculable, but the function $\tilde{K}$ at large
values of $b_T$ is nonperturbative.  We follow Ref.\ \cite{CSS1} to
separate the perturbative and nonperturbative parts of $\tilde{K}$.
First, we define
\begin{equation}
  \label{eq:bstar.def}
  {\bf b}_{*} = \frac{ {\bf b}_T }
                     { \sqrt{ 1 + b_T^2/b_{\rm max}^2} },
\qquad
  \mu_b = \frac{ C_1 }{ b_{*} }.
\end{equation}
Here $C_1$ is a fixed numerical coefficient and $b_{\rm max}$ is
chosen to keep $b_{*}$ in the perturbative region.  In the fits to
unpolarized Drell-Yan, the values chosen were
$b_{\rm  max}=0.5\,{\rm GeV^{-1}}$ in \cite{Landry:2002ix}, and
$b_{\rm  max}=1.5\,{\rm GeV^{-1}}$ in \cite{Konychev:2005iy}.
Next we write
\begin{equation}
  \tilde{K}(b_T;\mu)
  = 
  \tilde{K}(b_{*};\mu_b)
  - \int_{\mu_b}^\mu \frac{d\mu'}{\mu'} \gamma_K(g(\mu'))
  -g_K(b_T).
\end{equation}
The first two terms are perturbative and include all the evolution of
$\tilde{K}$. The last term is nonperturbative but scale independent.
It represents the only nonperturbative information needed to evolve
the Sivers function from the scale $Q_0$ where it was initially fit.
But this function is process independent \cite{collins}, so we can
take its value from already existing fits to unpolarized Drell-Yan 
\cite{Landry:2002ix,Konychev:2005iy} scattering at a variety of
energies.

This gives the evolved function:
\begin{widetext}
\begin{multline}
\label{eq:evolvedsiv1}
   \tilde{F}^{\prime \, \perp \, f}_{1T}(x,b_T;\mu,\zeta_F) 
=    \tilde{F}^{\prime \, \perp \, f}_{1T}(x,b_T;\mu_0,Q_0^2) 
\exp \Biggl\{ 
         \ln \frac{\sqrt{\zeta_F}}{Q_0} \tilde{K}(b_{\ast};\mu_b) 
         + \int_{\mu_0}^\mu \frac{d \mu^\prime}{\mu^\prime} \left[ \gamma_F(g(\mu^\prime);1) 
                 - \ln \frac{\sqrt{\zeta_F}}{\mu^\prime} \gamma_K(g(\mu^\prime)) \right] 
\\
         + \int_{\mu_0}^{\mu_b} \frac{d \mu^\prime}{\mu^\prime} \ln \frac{\sqrt{\zeta_F}}{Q_0}
                       \gamma_K(g(\mu^\prime))
         - g_K(b_T) \ln \frac{\sqrt{\zeta_F}}{Q_0}
\Biggr\}.
\end{multline}
\end{widetext}
We can set $\mu_0=Q_0$ and then use $Q_0=\sqrt{2.4}\,\mbox{GeV}$, which
is the appropriate scale for the fits in
\cite{Collins:2005ie,Anselmino:2008sga}, which used data from the
HERMES experiment.  For the prediction of data at a higher energy, one
should set $\mu^2=\zeta_F=Q^2$.  The anomalous dimensions $\gamma_F$ and $\gamma_K$
are used in a region where perturbative calculations are appropriate.

The Sivers function in transverse-momentum space is then obtained from
Eq.~(\ref{eq:evolvedsiv1}) by Fourier transformation, as
in Eq.~(\ref{eq:azimdep7}).

The one-loop values of the relevant perturbative quantities are listed
in 
the Appendix.

The size of the Sivers asymmetry is also often parametrized by the function
\begin{multline}
\label{eq:asymetry}
F_{f/P^{\uparrow}}(x,\trans{k};S,\mu,\zeta_F) - F_{f/P^{\uparrow}}(x,\trans{k};-S,\mu,\zeta_F) \\
= \Delta^N F_{f/P^{\uparrow}}(x,k_T;\mu,\zeta_F) \frac{\epsilon_{ij}  k_T^i S_T^j}{k_T} ,
\end{multline}
where
\begin{equation}
\label{eq:asymmetry2}
\Delta^N F_{f/P^{\uparrow}}(x,k_T) = -\frac{2 k_T}{M_p} F^{\perp \,f}_{1T}(x,k_T;\mu,\zeta_F).
\end{equation}

As can be seen from Figs.\ \ref{fig:siversevolved} and
\ref{fig:gaussfits} below, TMD functions broaden substantially as the
scale increases.  Thus larger values of transverse momentum become
important, and correspondingly we need the $\tilde{F}$ factor at small $b_T$.  

\subsection{Including the perturbative calculation of Sivers function at small-$b_T$}

At low scales, the Sivers function is dominantly at low values of
$k_T$, and correspondingly the range of $b_T$ that matters concerns
the larger values where both the starting value $\tilde{F}^{\prime \, \perp \,
  f}_{1T}(x,b_T;\mu_0,Q_0^2)$ and 
the evolution kernel $\tilde{K}(b_T;\mu)$ are in
the nonperturbative region.  After evolution to a sufficiently large
scale, the broadening of the $k_T$ distribution makes smaller values
of $b_T$ important, where there is perturbative information.  For both
this case and the treatment of the large-$k_T$ tail of the Sivers
function we can use the expansion (\ref{eq:small.b}) to write it in
terms of the twist-3 Qiu-Sterman function.  

\begin{widetext}
Following the method used for the unpolarized TMD PDF --- see
Ref.~\cite{CSS1,collins} and Eq.~(31) of Ref.~\cite{Aybat:2011zv} ---
we write
\begin{multline}
\label{eq:evolvedsiv2}
\tilde{F}^{\prime \, \perp \, f}_{1T}(x,b_T;\mu,\zeta_F) = 
 \sum_j \frac{M_p b_T}{2} \int_x^1 \frac{d \hat{x}_1 \, d \hat{x}_2}{\hat{x}_1 \, \hat{x}_2} 
   \tilde{C}^{\text{Sivers}}_{f/j}(\hat{x}_1,\hat{x}_2,b_{\ast};\mu_b^2,\mu_b,g(\mu_b)) \, 
   T_{F \,   j / P}(\hat{x}_1,\hat{x}_2,\mu_b) 
\\ 
\times \exp \left\{ 
     \ln \frac{\sqrt{\zeta_F}}{\mu_b} \tilde{K}(b_{\ast};\mu_b) + 
      \int_{\mu_b}^\mu \frac{d \mu^\prime}{\mu^\prime} \left[ \gamma_F(g(\mu^\prime);1) 
        - \ln \frac{\sqrt{\zeta_F}}{\mu^\prime} \gamma_K(g(\mu^\prime)) \right]
\right\} 
\times \exp \left\{ 
  -g^{\text{Sivers}}_{f/P}(x,b_T) - g_K(b_T) \ln \frac{\sqrt{\zeta_F}}{Q_0} 
\right\} .
\end{multline}
\end{widetext}
The first line describes the matching to a collinear treatment
relevant to small $b_T$.  There, $\tilde{F}^{'\,\perp \,
  f}_{1T}(x,b_T;\mu,\zeta_F)$ is expressed as a coefficient function
$\tilde{C}_{f/j}(\hat{x}_1,\hat{x}_2,b_{\ast};\mu_b^2,\mu_b,g(\mu_b))$
convoluted with a (twist-3) Qiu-Sterman function $T_{F \, j /
  P}(\hat{x}_1,\hat{x}_2,\mu_b)$, where for the simplicity, we neglected the terms proportional to the derivative of the twist-3 Qiu-Sterman function.  
On the second line, the first exponential comes from the perturbative
part of the evolution of the Sivers function; the use of $b_{*}$ and
$\mu_b$ ensures that $\tilde{C}$, $\tilde{K}$, $\gamma_F$, and $\gamma_K$ are
in the perturbative region.  The second exponential gives a
correction to allow for nonperturbative behavior at larger $b_T$.  In
its exponent are both the nonperturbative term $g_K(b_T)$ for the
evolution kernel, and an extra term $g^{\text{Sivers}}_{f/P}(x,b_T)$
for the Sivers function itself.  These terms are both
scale independent.~\footnote{Note that our sign convention on $g^{\text{Sivers}}_{f/P}(x,b_T)$ 
and $g_K(b_T)$ is opposite of Ref.~\cite{Aybat:2011zv} .}

The coefficient $\tilde{C}$ can be determined, for example, by
performing a low-order perturbative calculation of the left-hand side
of Eq.\ (\ref{eq:evolvedsiv2}), of the Qiu-Sterman function, and of the
first exponential, while ignoring the nonperturbative correction~\cite{Kang:2011mr}. 
The normalization factor, $M_p b_T/2$, in Eq.~\eqref{eq:evolvedsiv2} ensures that 
$T_F(\hat{x}_1,\hat{x}_2,\mu_b)$ has the standard normalization~\cite{Kang:2011mr}, and at zeroth order 
the contribution to the hard coefficient $\tilde{C}^{\text{Sivers}}_{f/j}(\hat{x}_1,\hat{x}_2,b_{\ast};\mu_b^2,\mu_b,g(\mu_b))$ is 
\begin{multline}
\label{eq:coeff0}
\tilde{C}^{\text{Sivers}, \; (0)}_{f/j}(\hat{x}_1,\hat{x}_2,b_{\ast};\mu_b^2,\mu_b,g(\mu_b)) = \\ \delta_{f,j} \, \delta(1 - x/\hat{x}_1) \, \delta(1 - x/\hat{x}_2),
\end{multline}
which is similar to the zeroth order term in Eq.~(A11) of Ref.~\cite{Aybat:2011zv} for the unpolarized case.  
(Recall that, since the Qiu-Sterman function $T_F(\hat{x}_1,\hat{x}_2,\mu_b)$ is universal, 
an extra minus sign is needed if we consider Drell-Yan instead of SIDIS.) 
The factor of $b_T$ in the normalization is a reminder that it is the \emph{derivative} 
of the Sivers function that we evolve in Eq.~\eqref{eq:evolvedsiv2}, not the Sivers function itself.
Higher-order contributions to the coefficient function 
can be taken directly from work, such as Ref.~\cite{Kang:2011mr}, which treats smaller $b_T$ within the Qiu-Sterman method.
Calculations of the 
unpolarized
coefficient functions to higher orders in the \MSbar{} scheme have already been carried out in 
Ref.~\cite{collins,Aybat:2011zv}.

The corresponding formula for the unpolarized TMD PDFs is very useful,
since instead of the Qiu-Sterman function it uses the ordinary
integrated PDFs, which are very well measured.  In contrast, the
phenomenology of the Qiu-Sterman function is less well known
quantitatively, so there may be less of an advantage of using Eq.\
(\ref{eq:evolvedsiv2}) instead of Eq.\ (\ref{eq:evolvedsiv1}).

In the remaining sections, we will discuss the implementation of evolution, given some nonperturbative
input functions, and 
provide specific evolved fits. 
Before continuing, however,
we should emphasize that matters related to the fitting of the nonperturbative
functions, including the choice of functional form for $g_K(b)$ and 
the matching procedure in Eq.~\eqref{eq:bstar.def}, are unrelated to the validity of the 
TMD-factorization formalism itself.
The TMD-factorization formalism automatically accommodates any refinements to 
knowledge about the nonperturbative physics.  Indeed, a central aim of this article is 
to demonstrate the generality of the method.  In our calculations below,
we have chosen to consider fits to the nonperturbative functions that correspond to detailed studies of existing data. 
In addition to providing tools for phenomenology, our calculations illustrate how 
numerical values for the Sivers function corresponding to the definition in 
Eq.~\eqref{eq:def2} can be obtained, once the nonperturbative functions are constrained by data.
Thus, our use of TMD-factorization is closely analogous to what already exists for collinear factorization.


\section{Gaussian Parametrizations in the Low-$q_T$ Region}
\label{sec:gaussian}
In this section we explain the implementation of QCD evolution for the Sivers function with a Gaussian ansatz. 
Since the small-$b_T$ region 
is twist-3, the tail of the (momentum-space) Sivers function (at
large $k_T$)
is power suppressed relative to the unpolarized TMD function. 
Furthermore, as illustrated in Ref.~\cite{Aybat:2011zv}, a Gaussian parametrization provides a 
good description of the low transverse-momentum behavior, even up to transverse momenta of a few GeV.
Therefore, we take as a starting point a detailed treatment of the twist-2 large-$b_T$ behavior, leaving for 
future refinements an account of the matching of the small-$b_T$
behavior to the twist-3 factorization formalism.     
That is, we use Eq.~\eqref{eq:evolvedsiv1} rather than
Eq.~\eqref{eq:evolvedsiv2}

Even so, a full treatment 
that extends to small-$b_T$ by including higher 
orders in $\tilde{C}_{f/j}(\hat{x}_1,\hat{x}_2,b_{\ast};\mu_b^2,\mu_b,g(\mu_b))$ will
be crucial in the long run for a complete understanding of 
the evolved Sivers function over the full range of $b_T$.  This is especially 
important to keep in mind when dealing with weighted integrals of the
Sivers function where the effect of the large transverse-momentum  
tail becomes magnified.  We intend to pursue this in future
refinements of the TMD approach.  

At the initial fitting scale, we drop the explicit scale dependence:
\begin{equation}
  \label{eq:F0.def}
  \tilde{F}^{\prime \, \perp }_{1T, \, 0}(x,b_T) = \tilde{F}^{\prime \, \perp \, f}_{1T}(x,b_T;\mu_0,Q_0^2). 
\end{equation}
To match previous fits \cite{Anselmino:2008sga,Collins:2005ie}, we
approximate the input function by a Gaussian
\begin{equation}
\label{eq:inprime}
\tilde{F}^{\prime \, \perp \, f}_{1T, \, 0}(x,b_T) 
= - \frac{\langle k_T^2 \rangle_0 f^{\perp}_{1T}(x) b_T}{2} \exp \left[- \langle k_T^2 \rangle_0 b_T^2 / 4 \right],
\end{equation}
which corresponds also to a Gaussian ansatz for the momentum-space distribution:
\begin{equation}
\label{eq:input}
F^{\perp }_{1T, \, 0 \, f}(x,k_T) = \frac{f^{\perp \, f}_{1T}(x)}{\langle k_T^2 \rangle_0^f \pi}  
\exp \left[ -k_T^2 / \langle k_T^2 \rangle_0^f \right].
\end{equation}
The parameter \kzero is the width of the Sivers function for a quark of flavor $f$ at the scale where the 
Gaussian fit is performed.  
Comparing with Eq.~\eqref{eq:evolvedsiv2}, we see that $g^{\rm Sivers}_{f/P}(x,b_T) = \langle k_T^2 \rangle_0^f b_T^2/4$.
The fits performed in~\cite{Anselmino:2008sga,Collins:2005ie} are for quite low scales 
($Q^2 = 2.4$~GeV$^2$ for HERMES data).  We therefore assume that the Sivers function  
is dominated by the nonperturbative large-$b_T$ region, in which case a Gaussian description, with 
a negligible tail effect, makes sense.
The first moment of the input momentum-space Sivers function obeys the usual relation:
\begin{equation}
\label{eq:firstmom}
f^{\perp \, (1)}_{1T, \, 0}(x) 
= \int d^2 \trans{k} \, \frac{k_T^2}{2 M_p^2} F^{\perp }_{1T, \, 0}(x,k_T) = \frac{\langle k_T^2 \rangle_0}{2 M_p^2} f^{\perp}_{1T}(x). 
\end{equation} 
We again remind the reader that for our calculations, we are assuming a Sivers function 
for SIDIS and that a sign flip is necessary to go to DY.
 
With $g_K(b_T)$ already known from previous fits to high energy 
Drell-Yan data~\cite{Ladinsky:1993zn,Landry:2002ix,Konychev:2005iy}, 
all that is now needed in order to obtain evolved Gaussian fits  
are $\langle k_T^2 \rangle_0^f$ and $f^{\perp \, f}_{1T}(x)$.  These will 
come from previously obtained fixed-scale Gaussian fits.  In the next 
section, we will provide two examples 
and illustrate the effect of evolution for two of the sets of Gaussian 
fits available in the literature.  

The function $g_K(b_T)$ 
is the only nonperturbative input that is necessary apart from these initial fits.  
We have also adopted the standard Gaussian ansatz for $g_K(b_T)$, writing 
$g_K(b_T)=g_2 b_T^2/2$.
Fits like those of Refs.~\cite{Ladinsky:1993zn,Landry:2002ix,Konychev:2005iy} provide 
numerical values for $g_2$.
In the Brock-Landry-Nadolsky-Yuan fits~\cite{Landry:2002ix} 
a value of $g_2 = 0.68$~GeV$^2$ is found.  This corresponds to a value for $b_{\rm max}$ of $0.5$~GeV$^{-1}$, and 
is what we will use in the fits of the next section.

\section{Specific Fits}
\label{sec:specificfits}

In this section 
we
provide examples of evolved fits, obtained 
by following the steps of Sec.~\ref{sec:gaussian} with specific 
fits for the input distributions.
We remind the reader that our numerical calculations correspond to the 
Sivers function of SIDIS, and that they acquire an overall minus sign in the Drell-Yan process.

\subsection{Bochum Fits}
\label{sec:bochum}
The fits of Ref.~\cite{Collins:2005ie} use a Gaussian 
to describe the HERMES measurements~\cite{Airapetian:2004tw} which 
were performed with  
an average $Q^2$ of $2.41$~GeV$^2$.
We refer to these as the \emph{Bochum fits}.
The function corresponding to $f^{\perp \, f}_{1T}(x)$ in Eq.~\eqref{eq:input}
is 
\begin{equation}
\label{eq:fxbochum}
\left[ f^{\perp \, {\rm up}/ {\rm down}}_{1T}(x) \right]_{\rm Bochum}
= \pm \frac{2 M_P^2}{\langle k_T^2 \rangle_0} A x^{b-1}(1-x)^5\,.
\end{equation}
The fit parameters are $$A = 0.17, \qquad b = 0.66.$$
In the Bochum fits, the parameter corresponding 
to \kzero{} in Eq.~\eqref{eq:input} is assumed to be independent 
of flavor and lies between $0.10$ and $0.32$~GeV$^2$.
We take 
\begin{equation}
\left. \kzero \right._{\rm Bochum}= \langle k_T^2 \rangle_{0 \, {\rm Bochum}} = 0.2 \, {\rm GeV}^2, 
\end{equation}
which corresponds to the ``best fit" scenario of
Ref.~\cite{Collins:2005ie}.

\begin{figure*}
\centering
\includegraphics[scale=.5]{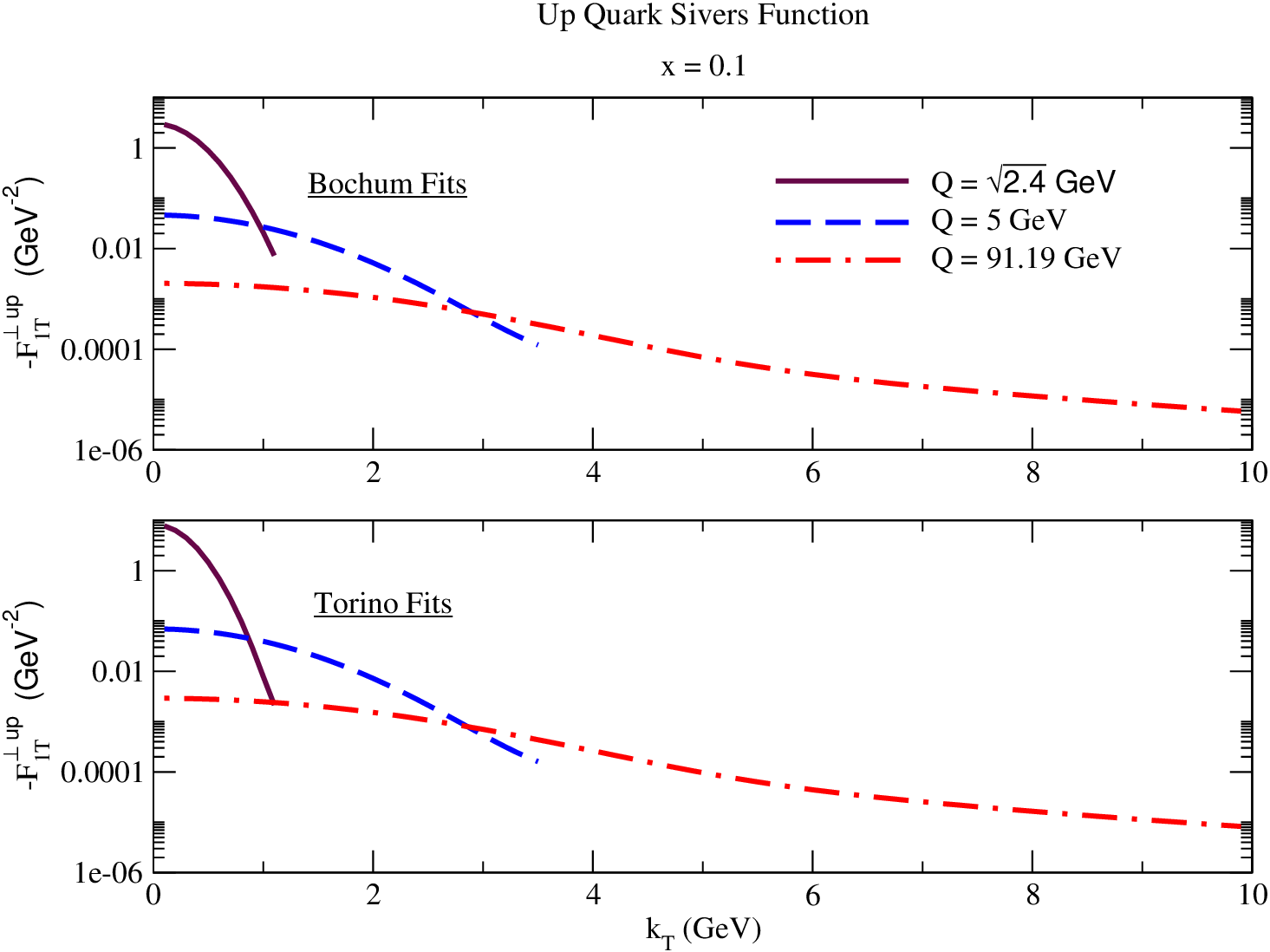}
\caption{(Color online.)
The (negative of the) up quark Sivers function 
at $x=0.1$ evolved from $Q=\sqrt{2.4}$~GeV(solid maroon) to
$Q=5$~GeV(dashed blue) 
and $Q=91.19$~GeV(dot-dashed red).  The upper plot is found by 
evolving the 
Gaussian fits of the Bochum group~\cite{Collins:2005ie}
and the lower plot is found 
by evolving the 
Gaussian fits of the Torino group~\cite{Anselmino:2008sga}.
In the case of the 
Bochum fits, the down quark Sivers function is just the negative of the up quark one.  
For the Torino fits, the down quark Sivers function is obtained by multiplying the up quark Sivers function by $-1.35$.
These functions acquire an overall reversal of sign if used in Drell-Yan.
}
\label{fig:siversevolved}
\end{figure*}

Samples of the result of using 
the Bochum fits in Eq.~\eqref{eq:evolvedsiv1} to evolve to different $Q$ are shown
in the 
upper
panel of Fig.~\ref{fig:siversevolved}.  The curves are shown for
$Q = \sqrt{2.4},5,91.19$~GeV since these are also the values already used
to illustrate the evolution of the unpolarized distribution functions in Ref.~\cite{Aybat:2011zv}.

\subsection{Torino Fits}
\label{sec:torino}
Next we consider the fits of Ref.~\cite{Anselmino:2008sga} which incorporated 
data from both HERMES~\cite{Diefenthaler:2007rj} and COMPASS~\cite{Martin:2007au,:2008dn}.
Again, the scale for the initial distributions is $Q^2 = 2.4$~GeV$^2$. 
We refer to these as the \emph{Torino fits}.
The function corresponding to $f^{\perp \, f}_{1T}(x)$ in Eq.~\eqref{eq:input} is 
\begin{equation}
\label{eq:fxtorino}
\left[ f^{\perp \, f}_{1T}(x) \right]_{\rm Torino}
= - \frac{M_p \sqrt{2 e} }{M_1 \langle k_T^2 \rangle} \mathcal{N}_f(x)\, f_f(x) \langle k_T^2 \rangle_0,  
\end{equation}
where
\begin{equation}
\mathcal{N}_f(x) \equiv N_f\,x^{\alpha_f}(1-x)^{\beta_f}\frac{(\alpha_f + \beta_f)^{(\alpha_f + \beta_f)}}{\alpha_f^{\alpha_f}\beta_f^{\beta_f}}\,,
\end{equation}
and $f_f(x)$ is the unpolarized parton distribution function for quarks of flavor $f$.
The fit parameters $N_f$, $\alpha_f$, $\beta_f$ are
\begin{align}
N_{\rm u} & = 0.35, \qquad \alpha_{\rm u} = 0.73, \qquad \beta_{\rm u} = 3.46, \\
N_{\rm d} & = 0.90, \qquad \alpha_{\rm d} = 1.08, \qquad \beta_{\rm d} = 3.46,
\end{align}
and $M_1^2 = 0.34$~GeV$^2$, $\langle k_T^2 \rangle = 0.25$~GeV$^2$.
The Gaussian slope parameter of the initial input distribution in the 
Torino fits is again flavor-independent and is
\begin{equation}
\label{eq:k0torino}
\left. \kzero \right._{\rm Torino} = \langle k_T^2 \rangle_{0 \, {\rm Torino}} 
= \frac{M_1^2 \langle k_T^2 \rangle}{M_1^2 + \langle k_T^2 \rangle}.
\end{equation}
For the integrated PDFs in Eq.~\eqref{eq:fxtorino}, we have used the 
lowest-order MSTW
parametrizations~\cite{Martin:2009iq,Martin:2009bu,Martin:2010db,mstwwebpage}.
Samples of the evolved
Torino fits are shown in the
lower 
panel of Fig.~\ref{fig:siversevolved}.

Note that there is over a factor of 2 difference between the Torino and the Bochum fits, and this 
gives a rough indication of the uncertainty involved in current treatments.
A discussion of the difference in the two methods can be found in Ref.~\cite{Anselmino:2005an}.

We do hope for future improvements of the fits.
A very recent parametrization of the nonperturbative input was presented in Ref.~\cite{Bacchetta:2011gx}.
The results are similar to the Torino fits above, but utilize a relation to generalized parton distributions, and allow
for a connection to a quantification of parton angular momentum.  Morever, model calculations, such as in Refs.~\cite{Courtoy:2008dn,Courtoy:2008vi}, and lattice QCD
calculations~\cite{Musch:2010ka} can aid in providing 
meaningful parametrizations of the nonperturbative input over the whole  of phase space and open up interesting questions regarding the matching of 
purely nonperturbative descriptions of the Sivers function to pQCD.

\begin{table*}
\caption{Table of evolved Gaussian parameters, obtained by fitting Gaussians to the 
evolved Bochum and Torino fixed-scale fits.  
The fits are for $x \Delta^N F_{f/P^{\uparrow}}(x,k_T;\mu,\zeta_F)$ and are related to $F^{\perp \,f}_{1T}(x,k_T;\mu,\zeta_F)$ via Eq.~\eqref{eq:asymmetry2}.
The parameters are listed for the up quark distributions at $x = 0.1$; the Sivers function
at different values of $x$ can be found by multiplying by the appropriate ratios obtained from 
Eqs.~(\ref{eq:fxbochum},~\ref{eq:fxtorino}).  
The Gaussian slope 
parameter $b^{\rm fit}$ is the same for the up and down quarks.  The normalization 
parameters $a_{\rm up}^{\rm fit}$ are related to the down quark 
normalizations by $a_{\rm down}^{\rm Bochum} = -a_{\rm up}^{\rm Bochum}$ and $a_{\rm down}^{\rm Torino} \approx -1.35 a_{\rm up}^{\rm Torino}$.  
The last two 
columns, $k_{T,{\rm max}}^{\rm Bochum}$ and $k_{T,{\rm max}}^{\rm Torino}$, are the values of $k_T$ above which the Gaussian fits drop to less than a ratio of $0.8$ of the Sivers functions 
calculated directly from Eq.~\eqref{eq:evolvedsiv1}. }
\begin{tabular}{|c|r@{.}l|r@{.}l|r@{.}l|r@{.}l|r@{.}l|r@{.}l|}
	\hline \hline
\multicolumn{13}{|c|}
	{\rule[-3mm]{0mm}{8mm} $x \Delta^N F_{f/P}^{\rm fit}(x=0.1,k_T) = a_f^{\rm fit} k_T e^{-b^{\rm fit} k_T^2}$}  \\
\textbf{\rm Q  (GeV)}
	& \multicolumn{2}{c|}{$b^{\rm Bochum}$ (GeV$^{-2}$) } 
	& \multicolumn{2}{c|}{$b^{\rm Torino} $ (GeV$^{-2}$)} 
	& \multicolumn{2}{c|}{$a_{\rm up}^{\rm Bochum}$ (GeV$^{-3}$)} 
	& \multicolumn{2}{c|}{$a_{\rm up}^{\rm Torino}$ (GeV$^{-3}$)} 
	& \multicolumn{2}{c|}{$k_{T,{\rm max}}^{\rm Bochum}$ (GeV)}
	& \multicolumn{2}{c|}{$k_{T,{\rm max}}^{\rm Torino}$ (GeV)} 
\\  \hline \hline
$\sqrt{2.4}$  & 4&9999   & 6&9382   & 6&5570 $\times 10^{-1}$ &  1&7763  $\times 10^{0}$  & .&.  &  .&.  \\
2.0           & 1&8251   & 2&0329   & 9&5506 $\times 10^{-2}$ &  1&6661 $\times 10^{-1}$  & .&.  &  .&.  \\
2.5           & 1&1726   & 1&2552   & 4&1658 $\times 10^{-2}$ &  6&7105 $\times 10^{-2}$  & 2&36 &  2&29 \\
3.0           & 0&9067   & 0&9555   & 2&5716 $\times 10^{-2}$ &  4&0138 $\times 10^{-2}$  & 2&56 &  2&50 \\
3.5           & 0&7604   & 0&7945   & 1&8430 $\times 10^{-2}$ &  2&8276 $\times 10^{-2}$  & 2&70 &  2&65 \\
4.0           & 0&6668   & 0&6929   & 1&4329 $\times 10^{-2}$ &  2&1745 $\times 10^{-2}$  & 2&80 &  2&76 \\
4.5           & 0&6013   & 0&6225   & 1&1718 $\times 10^{-2}$ &  1&7649 $\times 10^{-2}$  & 2&89 &  2&85 \\ 
5.0           & 0&5526   & 0&5705   & 9&9179 $\times 10^{-3}$ &  1&4854 $\times 10^{-2}$  & 2&96 &  2&92 \\  
10.0          & 0&3562   & 0&3637   & 3&9881 $\times 10^{-3}$ &  5&8409 $\times 10^{-3}$  & 3&39 &  3&36 \\  
15.0          & 0&2941   & 0&2992   & 2&5477 $\times 10^{-3}$ &  3&7049 $\times 10^{-3}$  & 3&56 &  3&54 \\  
20.0          & 0&2612   &0&2653    & 1&8893 $\times 10^{-3}$ &  2&7372 $\times 10^{-3}$  & 3&67 &  3&65 \\  
25.0          & 0&2400   & 0&2435   & 1&5090 $\times 10^{-3}$ &  2&1810 $\times 10^{-3}$  & 3&75 &  3&73 \\  
30.0          & 0&2249   & 0&2280   & 1&2602 $\times 10^{-3}$ &  1&8182 $\times 10^{-3}$  & 3&81 &  3&79 \\  
35.0          & 0&2135   & 0&2163   & 1&0841 $\times 10^{-3}$ &  1&5621 $\times 10^{-3}$  & 3&86 &  3&84 \\  
40.0          & 0&2044   & 0&2070   & 9&5257 $\times 10^{-4}$ &  1&3712 $\times 10^{-3}$  & 3&90 &  3&88 \\  
45.0          & 0&1969   & 0&1993   & 8&5046 $\times 10^{-4}$ &  1&2232 $\times 10^{-3}$  & 3&94 &  3&92 \\  
50.0          & 0&1907   & 0&1929   & 7&6878 $\times 10^{-4}$ &  1&1049 $\times 10^{-3}$  & 3&97 &  3&95 \\  
55.0          & 0&1853   & 0&1874   & 7&0188 $\times 10^{-4}$ &  1&0081 $\times 10^{-3}$  & 3&99 &  3&98 \\  
60.0          & 0&1806   & 0&1826   & 6&4604 $\times 10^{-4}$ &  9&2744 $\times 10^{-4}$  & 4&02 &  4&00 \\  
65.0          & 0&1765   & 0&1784   & 5&9868 $\times 10^{-4}$ &  8&5906 $\times 10^{-4}$  & 4&04 &  4&02 \\
70.0          & 0&1728   & 0&1747   & 5&5800 $\times 10^{-4}$ &  8&0035 $\times 10^{-4}$  & 4&06 &  4&04 \\
75.0          & 0&1695   & 0&1713   & 5&2267 $\times 10^{-4}$ &  7&4937 $\times 10^{-4}$  & 4&08 &  4&06 \\
80.0          & 0&1665   & 0&1683   & 4&9164 $\times 10^{-4}$ &  7&0467 $\times 10^{-4}$  & 4&10 &  4&08 \\
85.0          & 0&1638   & 0&1655   & 4&6421 $\times 10^{-4}$ &  6&6514 $\times 10^{-4}$  & 4&11 &  4&09 \\
90.0          & 0&1613   & 0&1629   & 4&3976 $\times 10^{-4}$ &  6&2993 $\times 10^{-4}$  & 4&13 &  4&11 \\
\hline
\end{tabular}
\label{table:gaussparams}
\end{table*}

\subsection{Evolved Gaussian Parametrizations}
\label{sec:evolvedparams}
\begin{figure*}
\centering
  \begin{tabular}{c}
    \includegraphics[scale=0.5]{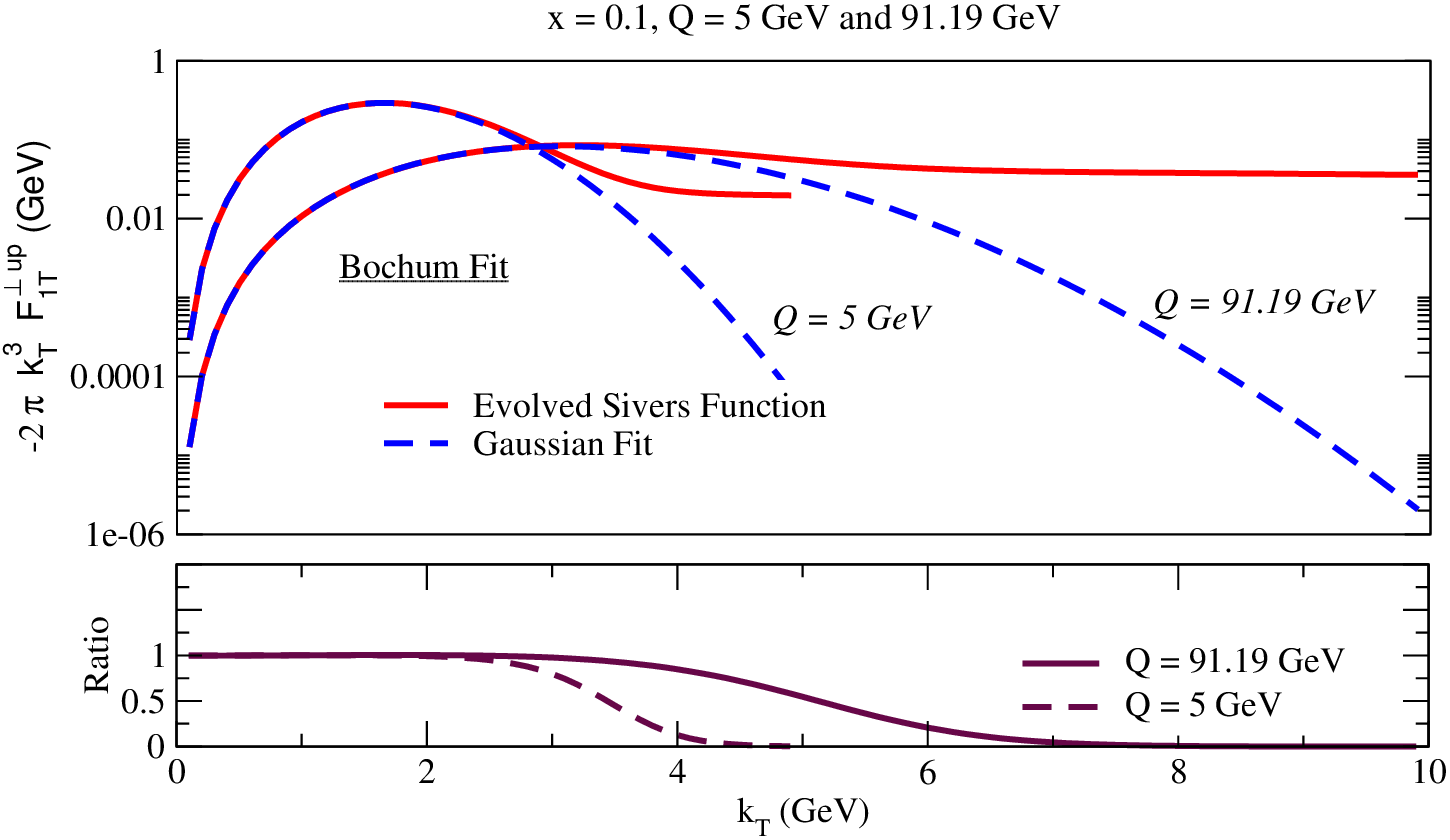} \\
    \vspace{6mm} \\
    \includegraphics[scale=0.5]{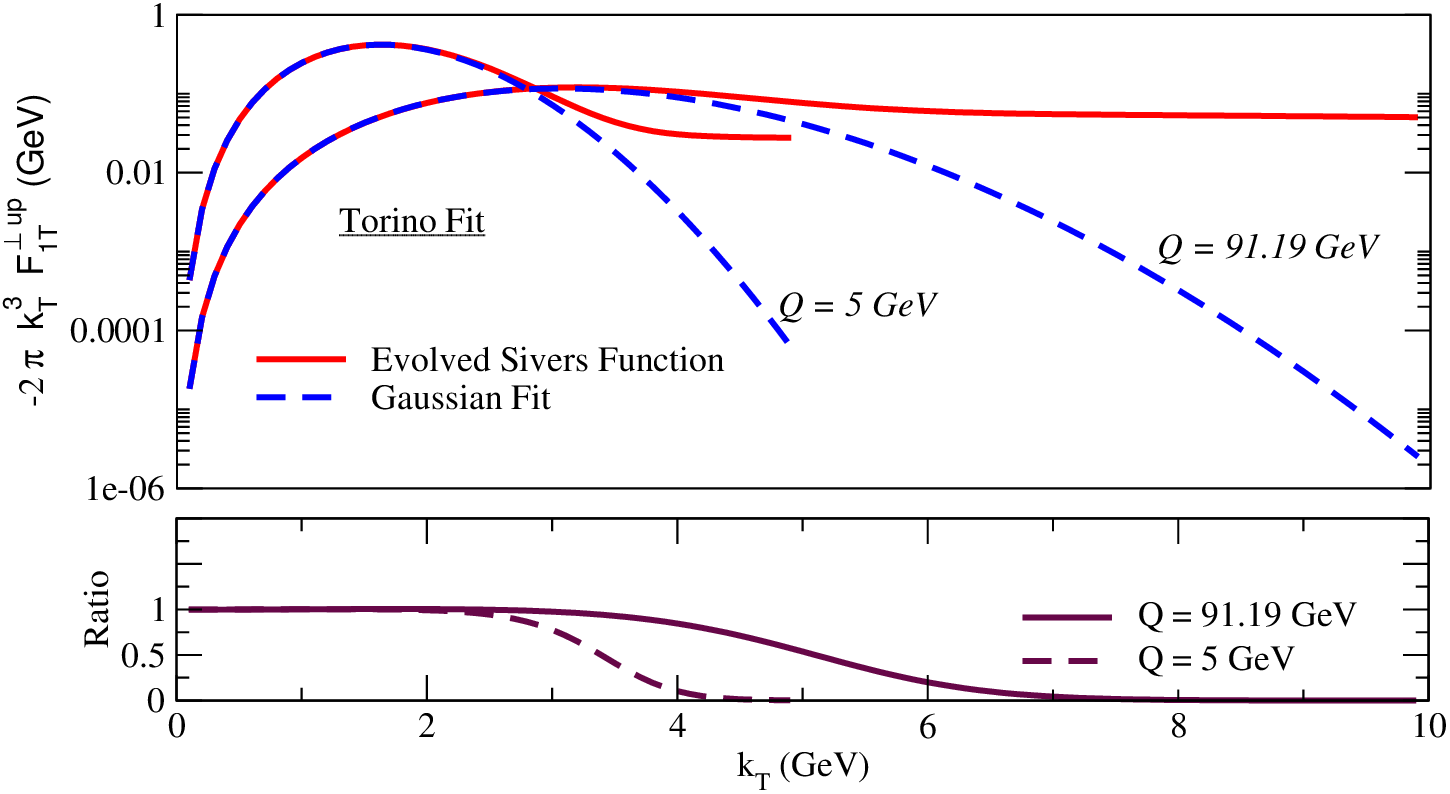}
  \end{tabular}
\caption{(Color online.)
The up quark Sivers function at $Q = 5$~GeV and $Q = 91.19$~GeV (solid curves) and the corresponding 
Gaussian fit for the low-$k_T$ region (dashed curves).  
Note that the function plotted is the Sivers function multiplied by
$-2\pi k_T^3$. 
The upper panel is obtained by 
evolving the 
Gaussian fits of the Bochum group~\cite{Collins:2005ie}
and lower panel is obtained by 
evolving the 
Gaussian fits of the Torino group~\cite{Anselmino:2008sga} .
Below each plot, the ratio
between a Gaussian fit and the evolved function including the tail is also shown.
}
\label{fig:gaussfits} 
\end{figure*}

Figure~\ref{fig:siversevolved} suggests that, apart from the tail at
large $k_T$, 
the Sivers function continues to be well described by a Gaussian shape, even after evolution 
to large $Q$.  To describe the evolution of a purely Gaussian parametrization, with the $x$ and 
$k_T$ dependence factorized, requires only 
a specification of the scale dependence of the Gaussian parameters.  
This saves having to directly calculate Eq.~\eqref{eq:evolvedsiv1}, and its 
transformation to momentum space, separately for 
each value of $Q$ and $x$.  Because of the general convenience 
of working with Gaussian functions, we have obtained Gaussian fits
for a range of $Q$ starting at $Q = \sqrt{2.4}$~GeV for the Bochum and Torino fits up to $Q = 90$~GeV.
The fits are obtained using the Wolfram \textsc{Mathematica 7 
FindFit} routine, and examples are shown as the dashed curves in Fig.~\ref{fig:gaussfits}. 
A table of the resulting values for the Gaussian parameters is shown
in Table~\ref{table:gaussparams}. 
(Fortran, C++, and Wolfram \textsc{Mathematica 7} code that produce evolved Gaussian fits is available at~\cite{webpage}.)
 
In Fig.~\ref{fig:gaussfits}, we illustrate the quality of the 
Gaussian fits to the Sivers function at intermediate and large $Q$ ($Q = 5$~GeV and $91.19$~GeV, respectively).  
In practice, the Sivers effect is often probed 
via observables like Eq.~\eqref{eq:firstmom}, so we have 
plotted the integrand, $-2 \pi k_T^3 F^{\perp \, {\rm up}}_{1T}(x,k_T;\mu,Q)$.  
Note that, after the evolution to large $Q$, the $-2 \pi k_T^3 F^{\perp \, {\rm up}}_{1T}(x,k_T;\mu,Q)$ acquires a very broad 
tail for both the Bochum and Torino fits.
The tail falls off slowly; for $Q=91.19$~GeV, the ratio of the value of the Bochum fit at $k_T = 10$~GeV to the 
value at $k_T = 5$~GeV is about $0.65$.  This is roughly consistent with the $1/ k_T$ fall-off at large $k_T$ that is expected
from the power counting arguments in Sec.~\ref{sec:powerlaws}.
The last two columns in Table~\ref{table:gaussparams} show the values of $k_T$ where the ratio
of the Gaussian fits to the original Sivers functions is $0.8$.  That is, above $k_{T,{\rm max}}^{\rm Torino}$ (GeV) the Gaussian fits 
to the evolved Torino Sivers function drop to less than $0.8$ of the original evolved Sivers function and similarly 
for $k_{T,{\rm max}}^{\rm Bochum}$.

That the description at small $k_T$ remains Gaussian is not entirely surprising given that the 
input we use for the nonperturbative evolution is Gaussian ($g_K(b_T) \propto b^2$).  However, it should 
be emphasized that the perturbative contribution to evolution results in
a substantial modification of the shape and normalization  
of the TMD PDF, even at low $k_T$.  Therefore, Table~\ref{table:gaussparams} is \emph{not} 
the result of simply Fourier transforming the nonperturbative 
contribution to Eq.~\eqref{eq:evolvedsiv1}.  Rather, to get the right TMD PDF, even when using a Gaussian approximation 
for low $k_T$, the full pQCD evolution must be included.
We find that difference between the fitted Gaussian and the result obtained by naively Fourier transforming 
the nonperturbative part of the evolution
is similar to what was found for the unpolarized TMD PDF (see Fig.\ 2 of Ref.~\cite{Aybat:2011zv}).   

The presence of the tail illustrates the danger in evaluating moment
integrals like Eq.~\eqref{eq:firstmom} without  
a careful account of the large-$k_T$ behavior. 
For $Q = 91.19$~GeV, there is more than 
40\% suppression in the integral of the curves in Fig.~\ref{fig:gaussfits} from $0$ to $10$~GeV when the 
Gaussian fit is used rather than the fit including the tail.  (Note that in principle the integral should be extended 
to order $Q$.)  For the $Q = 5$~GeV curves, integrated up to $5$~GeV, the corresponding suppression is only about 9\%.

\begin{figure*}
\centering
\includegraphics[scale=.6]{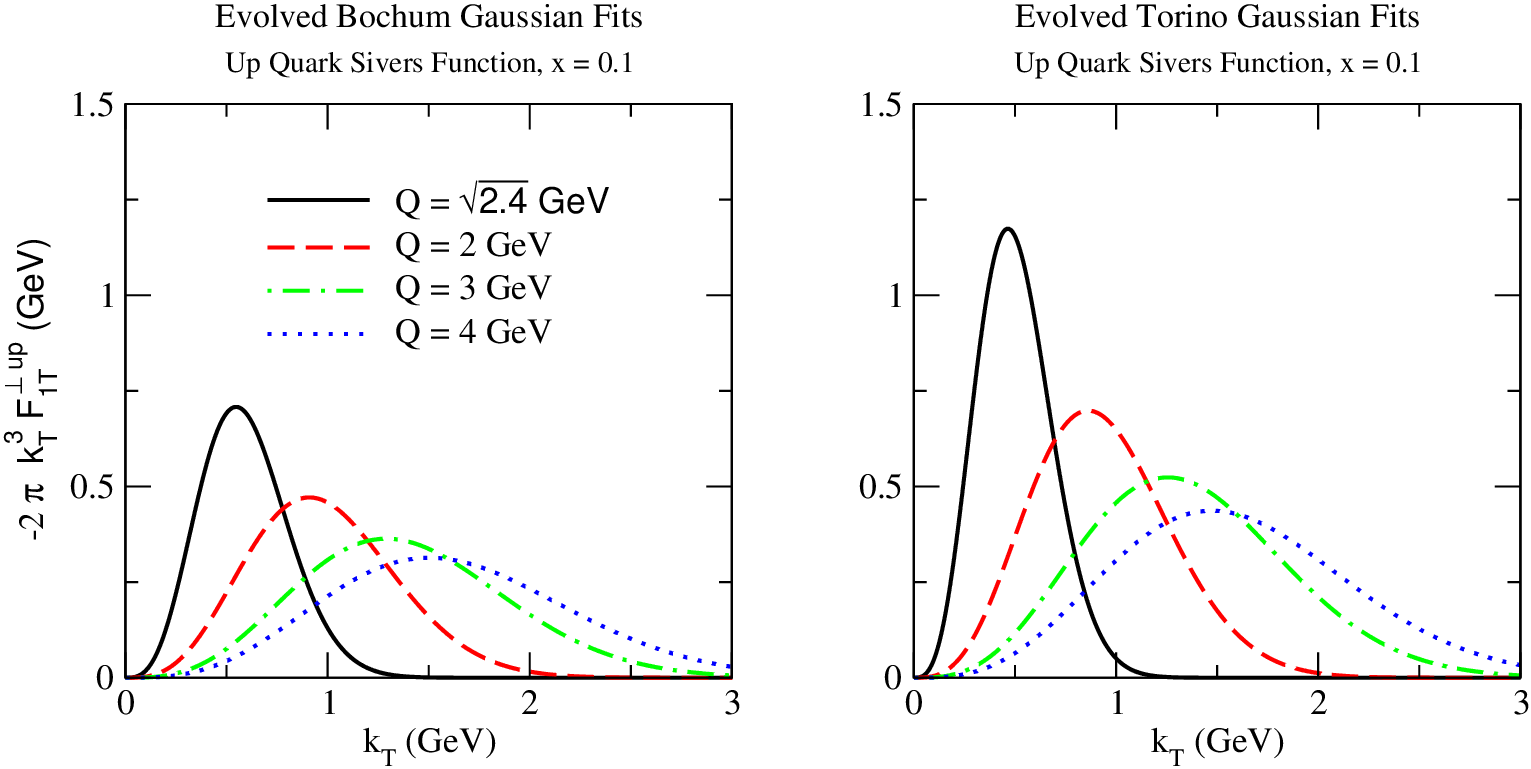}
\caption{(Color online.)
The evolving Gaussian parameters for $-2 \pi k_T^3 F^{\perp \, {\rm up}}_{1T}(x,k_T;\mu,Q)$ 
for a range of $Q$ obtained from the Torino and Bochum fits.  Table~\ref{table:gaussparams} lists the 
Gaussian parameters for a selection of $Q$. 
}
\label{fig:gaussfits2}
\end{figure*}

By contrast, at low-$k_T$ the Gaussian functions, shown as the
dashed curves in Fig.~\ref{fig:gaussfits}, provide excellent approximations to the evolved Sivers function.  
This suggests that the evolved Gaussian approximation is especially suited to low-$Q$/low-$k_T$ studies.
A sample of evolved Gaussian 
fits for lower $Q$ is shown in Fig.~\ref{fig:gaussfits2}.


\section{Discussion and Conclusions}
\label{sec:discussion}
Many of the recent phenomenological efforts related to the 
study of transverse polarization effects in TMDs have assumed 
a lowest-order, generalized parton-model (GPM) picture~\cite{D'Alesio:2007jt} 
and work within a rather narrow range of energy scales.  However, the 
full power of factorization theorems lies in their ability to make 
predictions for a variety of processes over a wide range of energy scales.  
In this article, we have explained the steps for implementing evolution 
for polarization dependent TMD PDFs, specifically illustrated with the 
Sivers function.  The basic method is the CSS formalism~\cite{Collins:1981va,CS1,CSS1}, 
with the specific formulation of the TMD-factorization formalism given recently in Ref.~\cite{collins}. 
An advantage of the most up-to-date TMD-factorization formula is 
that it is written in a 
form closely analogous to the GPM (see Eq.~\eqref{eq:parton2}), 
with explicit definitions for the individual TMDs.
Therefore, existing treatments that rely on a GPM framework need only 
to replace the unevolved TMDs with the evolved ones.
An important aspect of our approach is that it relies on a genuine, complete 
TMD-factorization formalism, to be contrasted with the resummation methodology
that has often been relied on in the past to treat many aspects of TMD physics.  That is, the TMD-factorization 
formalism provides, from the outset, a consistent treatment of factorization for the full range of $k_T$ (or, equivalently, 
the full range of $b_T$ in coordinate space). 

Fortunately, many of the results obtained from the treatment of unpolarized
TMDs can be carried over directly to the polarization dependent case, including 
the calculation of the anomalous dimensions $\gamma_F$, $\gamma_D$ and $\gamma_K$,
and the CS evolution kernel $K$, in both its calculable perturbative
part and its nonperturbative part $g_K(b_T)$ that is known from fits
to unpolarized Drell-Yan.  
An important difference from the unpolarized case is in 
the matching at large-$k_T$.  In the unpolarized case, the TMD PDF (or FF) matches 
to a twist-2 collinear factorization treatment at large $k_T$, whereas the Sivers function 
matches to a twist-3 collinear factorization treatment related to the Qiu-Sterman formalism, as 
in Eq.~\eqref{eq:evolvedsiv2}.  Thus, the treatment provided in this article unifies several different 
aspects of TMD physics.

It is worth commenting on the often repeated statement (see, e.g., Ref.~\cite{Cherednikov:2011ku}) that
calculations in covariant gauges are impractical or inconvenient, and that working 
in light-cone gauge is therefore preferred.
In our work, we find that the opposite is true.
Namely, the calculation of the perturbative parts (at least to order $\alpha_s$)  follows clear-cut steps in Feynman 
gauge, while the derivation of TMD-factorization theorems is much more direct in Feynman 
gauge than in light-cone gauge.  (Indeed, we are not aware of the
existence of a detailed light-cone gauge derivation of TMD factorization.) 
Moreover, once the calculation of the perturbative parts has been performed in 
Feynman gauge, a generalized 
parton-model interpretation follows directly from the TMD-factorization formula in Eq.~\eqref{eq:parton2}.
For these reasons, we advocate continuing to work in Feynman gauge for both calculations and derivations.

We have implemented the evolution explicitly using as input the already known 
$\gamma_F$, $\gamma_D$ and $\gamma_K$ (supplied for easy reference 
in 
the Appendix,
previous fixed-scale Gaussian fits of the Sivers function at 
low-$Q$~\cite{Collins:2005ie,Anselmino:2008sga}, and previous 
fits of the CSS formalism to DY~\cite{Landry:2002ix}.
For the explicit calculations in the present article, we have focused only on the low-$k_T$ region where we need not 
be concerned with the treatment of the Qiu-Sterman formalism at large $k_T$, and 
the approximations of Sec.~\ref{sec:gaussian} make sense.
The resulting evolved momentum-space Sivers functions are shown in 
Fig.~\ref{fig:siversevolved}.  Comparing with Fig.~1 of Ref.~\cite{Aybat:2011zv} for the
evolution of the unpolarized TMD PDF, one sees even more suppression 
as $Q$ is increased than in the unpolarized case.
Also note that a significant perturbative tail 
is generated at large $Q$ as shown in Fig.~\ref{fig:gaussfits}.  
We reemphasize that this should be
kept in mind when evaluating integrals like Eq.~\eqref{eq:firstmom}.

Gaussian parametrizations are particularly convenient for 
doing explicit calculations.  Therefore, we have tested the 
quality of Gaussian fits after evolution to large $Q$ and find 
that the Gaussian function provides an excellent approximation 
to the Sivers function at small $k_T$, even for $Q \approx 90.0$~GeV.  
We have made these fits available, as well
as code for generating evolved TMDs at a website maintained 
by two of us (Aybat and Rogers)~\cite{webpage}.

Much work remains to be done in the effort to connect a full QCD treatment 
of TMDs with phenomenology.  An explicit 
implementation of the  matching to the twist-3 Qiu-Sterman 
formalism is still needed, and will be particularly important 
for a correct treatment of $k_T$-weighted observables in which the 
extra $k_T$ factors enhance the contribution from the large 
$k_T$ region. 
The recent work of Ref.~\cite{Kang:2011mr} may help. 
Moreover, as new data become available for both polarized 
and unpolarized cross sections, it will be useful to construct new 
fits that include evolution from the beginning.    
Finally, explicit calculations, analogous to the ones presented here, need to be
applied to the other TMDs like the Boer-Mulders and Collins functions.

At large $Q$, the shape of the distribution is especially sensitive to the value of $b_{\rm max}$,
$g_2$ and the functional form of $g_K(b_T)$.  Reference \cite{Konychev:2005iy}, for example, finds that 
a larger value of $b_{\rm max}$ is preferred, along with a corresponding change in $g_2$.
Furthermore, Refs.~\cite{Qiu:2000hf,Berger:2002ut} find advantages to using a 
different functional form, $\sim b^{2/3}$ rather than $\sim b^2$, for $g_K(b_T)$.  
This should be taken into account in future improvements to the fits.  The 
particular set of parameters used in the calculations in the present article 
were chosen both because of their simplicity and 
because they correspond to the current state-of-the-art  of 
global fits to the unpolarized Drell-Yan cross section.

In the future, model calculations (see,e.g.,~\cite{Avakian:2009jt} and references therein for an overview)
can be potentially helpful for fixing nonperturbative input.  Certain models also lead to 
nonperturbative input distributions that deviate from the Gaussian ansatz.  Conversely, 
incorporating evolution into model calculations can help establish the scale appropriate to the model.

Theoretical uncertainties in the TMD fits, both for unpolarized 
and polarized TMDs, can be reduced by including higher-order results 
for the anomalous dimensions and the CSS kernel $K$ 
(in the perturbative region). 
Fortunately, as 
we have discussed in this paper, these anomalous dimensions and the 
kernel $K$ are the same for unpolarized TMDs and the Sivers function. 
Therefore by calculating them at next-to-next-to-leading order
in pQCD, we can reduce the 
theoretical uncertainties for both unpolarized and polarized TMDs at the same time. 

The ultimate goal 
is to obtain sets of TMD PDFs and FFs 
that can be used in a way that is closely analogous 
to what already exists for processes that use collinear factorization.  Namely, 
we would like to obtain a set of TMD fits based on 
precise TMD definitions such that they can be reliably used to make predictions.


\begin{acknowledgments} 
M.~Aybat and T.~Rogers acknowledge support from  the research 
program of the ``Stichting voor Fundamenteel Onderzoek der Materie (FOM),'' 
which is financially supported by the ``Nederlandse Organisatie voor
Wetenschappelijk Onderzoek (NWO).'' 
M.~Aybat also acknowledges support from the FP7 EU-programme
HadronPhysics2 (Contract No.\ 2866403).
T.~Rogers was also supported in part by the National Science
Foundation, Grant No.\ PHY-0969739.
J.~C.~Collins and J.~W.~Qiu were supported by the U.S. Department of 
Energy under Grant No.\ DE-FG02-90ER-40577 and Contract No.\
DE-AC02-98CH10886, respectively. 
M.~Aybat and T.~Rogers thank Christine Aidala, Aurore Courtoy, Bryan Field, and Alexei Prokudin  for useful discussions.
\end{acknowledgments}

\appendix

\section*{APPENDIX: ANOMALOUS DIMENSIONS ETC.}
Here we list the \MSbar{}-scheme anomalous dimensions~\cite{collins} that were 
used in, for example, Eqs.~\eqref{eq:evolvedsiv1} and \eqref{eq:evolvedsiv2}:
\begin{equation}
\label{eq:anom}
\gamma_{\rm F}(\mu;\zeta_F / \mu^2) = \alpha_s  \frac{C_{\rm F}}{\pi} \left(\frac{3}{2} - \ln \left( \frac{\zeta_F}{\mu^2} \right) \right)
+ \mathcal{O}(\alpha_s^2).
\end{equation}
At order $\alpha_s$, the quark TMD FF anomalous dimension is the same as for the TMD PDF.
The CS kernel up to order $\alpha_s$ in ${\bf b}_T$ space is
\begin{equation}
\label{eq:kern}
\tilde{K}(\mu,b_T) = - \frac{\alpha_s C_F}{\pi} \left[ \ln(\mu^2 b_T^2) - \ln 4 + 2 \gammae \right]
+ \mathcal{O}(\alpha_s^2).
\end{equation}
The anomalous dimension of $\tilde{K}$ is up to order $\alpha_s$,
\begin{equation}
\label{eq:Kanom}
\gamma_K(\mu) = 2 \frac{\alpha_s C_F}{\pi} + \mathcal{O}(\alpha_s^2).
\end{equation}

\bibliography{SiversFinal2-update}

\end{document}